\PassOptionsToPackage{font=small}{caption}

\documentclass[journal]{IEEEtran}
\usepackage[square,sort,comma,numbers]{natbib}
\usepackage{multirow}
\usepackage{subfig}
\usepackage{graphicx}

\newcounter{subcopyrightbox@save}
\usepackage{caption}
\usepackage{color, url}
\usepackage{xspace} 
\usepackage[ruled,linesnumbered]{algorithm2e}
\usepackage{mathrsfs}
\usepackage{amssymb}
\usepackage{amsmath}
\usepackage{epstopdf}

\newcommand{\argmin}{\operatornamewithlimits{argmin}}

\newcommand{\myparatight}[1]{\smallskip\noindent{\bf {#1}:}~}
\newenvironment{packeditemize}{\begin{list}{$\bullet$}{\setlength{\itemsep}{2pt}\addtolength{\labelwidth}{4pt}\setlength{\leftmargin}{20pt}\setlength{\listparindent}{\parindent}\setlength{\parsep}{0pt}\setlength{\topsep}{0pt}}}{\end{list}}

\begin{document}

\title{On the Security of Trustee-based Social Authentications}

\author{Neil Zhenqiang Gong,~\IEEEmembership{Student Member, IEEE} and Di Wang
\thanks{
N. Z. Gong and D. Wang are with the Electrical Engineering and Computer Science Department, University of California at Berkeley, Berkeley, CA, 94720 USA, e-mail: (neilz.gong@berkeley.edu; wangd@cs.berkeley.edu).}
\thanks{We would like to thank Dawn Song for insightful discussions. This work is supported by the NSF under Grants No. CCF-0424422, 0842695, 0831501 CT-L, by the AFOSR under MURI Award No. FA9550-09-1-0539, by the Office of Naval Research under MURI Grant No. N000140911081, and by Intel through the ISTC for Secure Computing. Any opinions, findings, and conclusions or recommendations expressed in this material are those of the author(s) and do not necessarily reflect the views of the funding agencies. 
}
}


\maketitle

\begin{abstract}

Recently, authenticating users with the help of their friends (i.e., \emph{trustee-based social authentication}) has been shown to be a promising \emph{backup} authentication mechanism~\cite{Brainard06,Schechter09,facebookSocAuth,facebookTF}. A user in this system is associated with a few \emph{trustees} that were selected from the user's friends. When the user wants to regain access to the account, the service provider sends \emph{different} verification codes to the user's trustees. The user must obtain at least $k$ (i.e., \emph{recovery threshold}) verification codes from the trustees before being directed to reset his or her password.

In this paper,  we provide the first systematic study about the security of trustee-based social authentications. Specifically,  we first introduce a novel framework of attacks, which we call \emph{forest fire attacks}. In these attacks, an attacker initially obtains a small number of compromised users, and then the attacker iteratively attacks the rest of users by exploiting trustee-based social authentications. 
Then, we construct a probabilistic model to formalize the threats of forest fire attacks and their costs for attackers. Moreover, we introduce various defense strategies.  Finally, we apply our framework to extensively evaluate  various concrete attack and  defense strategies using three real-world social network datasets. Our results have strong implications for the design of more secure  trustee-based social authentications.

\end{abstract}

\IEEEpeerreviewmaketitle

\section{Introduction}

Web services (e.g., Gmail, Facebook, and online Bankings) today most commonly rely on passwords to authenticate users. Unfortunately, two serious issues in this paradigm are: users will inevitably forget their passwords, and their passwords could be compromised and changed by attackers, which result in the failures to access their own accounts.  

Therefore,  web services often provide users with backup authentication mechanisms to help users regain access to their accounts. Unfortunately, current widely used backup authentication mechanisms such as  security questions and alternate email addresses are insecure or unreliable or both. Previous works~\cite{Poddand96, Schechter09-oakland, Zviran90} have shown that security questions are easily guessable and phished, and that users  might  forget their answers to the security questions. A previously registered alternate email address might expire upon the user's change of school or job. For the above reasons, it is important to design a secure and reliable backup authentication mechanism. 



Recently, \emph{trustee-based  social authentication} has attracted increasing attentions and has been shown to be a promising backup authentication mechanism~\cite{Brainard06,Schechter09,facebookSocAuth,facebookTF}. Brainard et al.~\cite{Brainard06} first proposed trustee-based social authentication and combined it with other authenticators (e.g., password, security token) as a two-factor authentication mechanism. Later, trustee-based social authentication was adapted to be a backup authenticator~\cite{Schechter09,facebookTF,facebookSocAuth}. In particular, Schechter et al.~\cite{Schechter09} designed and built a prototype of trusted-based social authentication system which was integrated into Microsoft's Windows Live ID. Schechter et al. found that trustee-based social authentication is highly \emph{reliable}. Moreover, Facebook announced its trustee-based social authentication system called \emph{Trusted Friends} in October, 2011~\cite{facebookTF}, and it was redesigned and improved to be \emph{Trusted Contacts}~\cite{facebookSocAuth} in May, 2013.


However, these previous work either focus on security at individual levels~\cite{Brainard06,Schechter09} or totally ignore security~\cite{facebookTF,facebookSocAuth}. In fact, security of users are correlated in trustee-based social authentications, in contrast to traditional authenticators (e.g., passwords, security questions, and fingerprint)  where security of  users are independent. Specifically, a user's security in trustee-based social authentications relies on the security of his or her trustees; if all trustees of a user are already compromised, then the attacker can also compromise him or her because the attacker can easily obtain the verification codes from the compromised trustees.  The impact of this key difference has not been touched. Moreover, none of the existing work has studied the fundamental design problems such as \emph{how to select trustees for users so that the system is more secure and how to set the system parameters (e.g., recovery threshold) to balance between security and usability}.

\myparatight{Our work} In this paper, we aim to provide the first systematic study about the security of trustee-based social authentications. 
To this end, we first propose a novel framework of attacks that are based on the observation that users' security are correlated in trustee-based social authentications. In these attacks, an attacker initially obtains a small number of compromised users which we call \emph{seed users}. The attacker then iteratively attacks other users according to some \emph{priority ordering} of them. In an \emph{attack trial} to a user Alice, if at least $k$ trustees of Alice are already compromised, then the attacker can easily compromise Alice; otherwise the attacker can (optionally) send spoofing messages to Alice's uncompromised trustees to request verification codes, and such spoofing attacks can succeed with some probability~\cite{Schechter09}. Our attacks are similar to forest fires which start from a few points and spread among the forests. Thus, we call them \emph{forest fire attacks}. 


Second, we construct a probabilistic model to formalize the threats of forest fire attacks and their costs for attackers. For each user, our model computes the \emph{compromise probability} that the user is compromised after a given number of attack iterations. With those compromise probabilities, our model calculates the \emph{expected number of compromised users} and treats it as the threat. Moreover, our model quantifies the costs of sending spoofing messages for attackers. 

Third,  we explore various scenarios where seed users have different properties and introduce strategies to  construct priority orderings. For instance, one scenario could be that  seed users happen to be appointed as trustees of a large number of users. Furthermore, we discuss a few defense strategies. For example, one strategy is to guarantee that no user is appointed as a trustee of a large number of users.



\myparatight{Results and impact of our work} 
We apply our framework to extensively evaluate various concrete attack scenarios, defense strategies, and the impact of system parameters using three real-world social networks. First, we find that forest fire attack is a potential big threat. In particular, when all the users with at least 10 friends in these social networks  adopt trustee-based social authentications,
 an attacker can compromise tens of thousands of users in some cases even if the number of seed users is 0; using a small number (e.g., 1,000) of seed users, the attacker can further compromise two to three orders of magnitude more users with low (or even no) costs of sending spoofing messages. Second, our defense strategy, which guarantees that no users are selected as trustees by too many other users, can decrease the expected number of compromised users by one to two orders of magnitude and increase the costs for attackers by a few times in some cases. Third, we find that, in contrast to existing work~\cite{Schechter09,facebookSocAuth,facebookTF} where the recover threshold is set to be three, it could be set to be four to better balance between security and usability. 

In summary, our key contributions are as follows:

\begin{packeditemize}
\item We propose a novel framework of attacks, which we call \emph{forest fire attacks}.

\item We construct a model to formalize the threats of forest fire attacks and their costs for attackers. Moreover, we explore various attack scenarios and defense strategies.  

\item We apply our framework to extensively evaluate these attack scenarios, defense strategies, and the impact of system parameters using three real-world social networks. Our results have strong implications for designing more secure trustee-based social authentications.

\end{packeditemize}

\section{Background}
First, we overview how a trustee-based social authentication system works. Then, we introduce two basic concepts, i.e., 
\emph{social networks} and  \emph{trustee networks}.

\subsection{Trustee-based social authentications}

A trustee-based social authentication includes two phases:
\begin{packeditemize}
\item {\bf Registration Phase.} The system prepares trustees for a user Alice in this phase. Specifically, Alice is first authenticated with her main authenticator (i.e., password), and then a few (e.g., 5) friends, who also have accounts in the system, are selected by either Alice herself or the service provider from Alice's  friend list and are appointed as Alice's trustees. 
\item {\bf Recovery Phase.} When Alice forgets her password or her password was compromised and changed by an attacker, she recovers her account with the help of her trustees in this phase. Specifically, Alice first sends an account recovery request with her \emph{username} to the service provider which then shows Alice an URL. Alice is required to share this URL with her trustees. Then, her trustees authenticate themselves into the system and retrieve verification codes using the given URL.  Alice then obtains the verification codes from her trustees via emailing them, calling them, or meeting them in person. If Alice obtains a sufficient number (e.g., 3) of verification codes and presents them to the service provider, then Alice is authenticated and is directed to reset her password. We call the number of verification codes required to be authenticated the \emph{recovery threshold}. 
\end{packeditemize}

Note that it is important for Alice to know who her trustees are in the Recovery Phase.  Schechter et al.~\cite{Schechter09} showed that users cannot remember their trustees via performing user studies. Thus, a usable trustee-based social authentication system should remind Alice of her trustees.

Next, we provide details about two representative trustee-based social authentication systems which were implemented by Microsoft~\cite{Schechter09} and Facebook~\cite{facebookSocAuth,facebookTF}, respectively.

\myparatight{Microsoft's trustee-based social authentication} Schechter et al.~\cite{Schechter09} designed and built a  trustee-based social authentication system and integrated it into Microsoft's Windows Live ID service. In the Registration Phase, users provide \emph{four} trustees. The recovery threshold is \emph{three}. Moreover, users will be reminded of their trustees. 

\myparatight{Facebook's trustee-based social authentication} Facebook's trustee-based social authentication system is called \emph{Trusted Friends}~\cite{facebookTF}, whose improved version is \emph{Trusted Contacts}~\cite{facebookSocAuth}. In the Registration Phase of Trusted Contacts, a user selects \emph{three to five} friends from his or her friend list as trustees. The recovery threshold is also set to be \emph{three}. Facebook does not remind a user of his or her trustees, but it asks the user to type in the names of his or her trustees instead. However, once the user gets one trustee correctly, Facebook will remind him or her of the remaining trustees. 

Both trustee-based social authentication systems ask users to select their own trustees without any constraint. In our experiments (i.e., Section~\ref{sec:eva}),  we show that the service provider can constrain trustee selections via imposing that no users are selected as trustees by too many other users, which can achieve better security guarantees. Moreover, none of these work performed rigorous studies to support the choice of three as the recovery threshold. In fact, our experimental results show that setting the  recovery threshold to be four could better balance between security and usability.

\begin{figure*}[!t]
\centering
\vspace{-4mm}
\includegraphics[width=0.95\textwidth, height=2in]{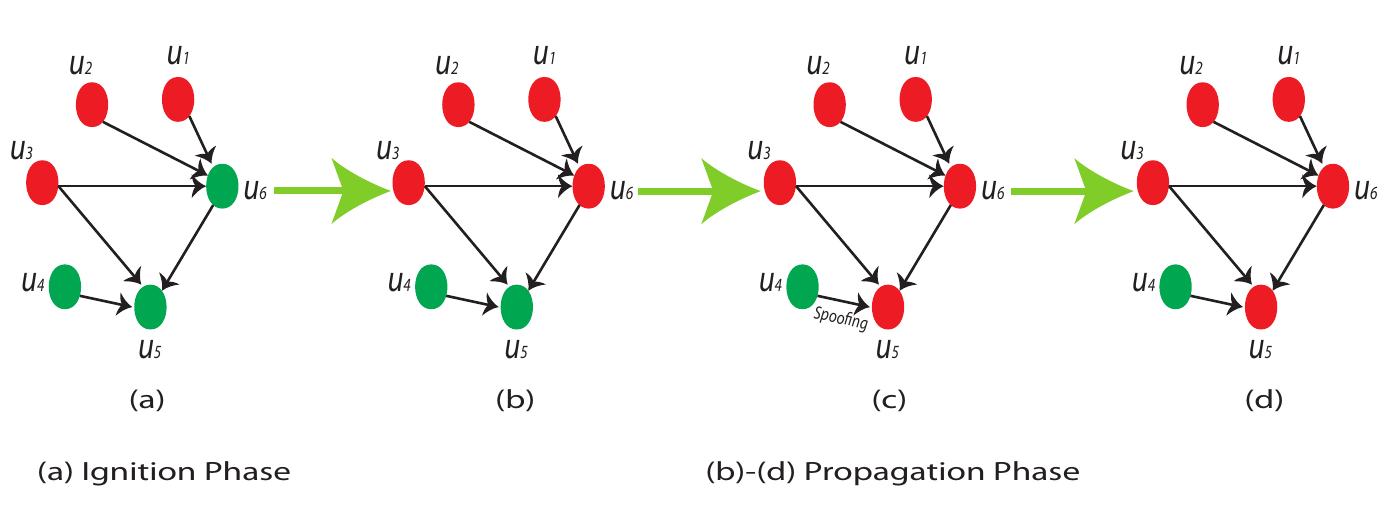}
\caption{Illustration of a forest fire attack to a service with 6 users. The shown graph is the trustee network. Recovery threshold is three. Users $u_5$ and $u_6$ have adopted the trustee-based social authentication.   The attack ordering is $u_6,u_5,u_4$. (a) $u_1, u_2$, and $u_3$ are compromised seed users. (b) $u_6$ is compromised because three of his or her trustees are already compromised. (c) $u_5$ is compromised because the attacker already compromises his or her trustees $u_3$ and $u_6$ and obtains a verification code from $u_4$ via spoofing attacks. (d) $u_4$ is not compromised because he or she hasn't adopted the service.}
\label{forest_fire_attack}
\vspace{-4mm}
\end{figure*}

\subsection{Social networks and trustee networks}
We denote a social network as $G=(V,E)$, where each node in $V$ corresponds to a user in the service and an undirected edge $(u,v)$ represents that users $u$ and $v$ are  friends. Moreover, in a trustee-based social authentication system,  users and their trustees  form a directed network. We call this directed network a \emph{trustee network} and denote it as $G_T=(V_T, E_T)$, where a node in $V_T$ is a user in the service and a directed edge $(v,u)$ in $E_T$ means $v$ is a trustee of $u$. 

One fundamental challenge in trustee-based social authentication is how to construct the trustee network from a social network so that the system is more secure.


%


\section{Threat Model}

We first introduce attackers' background knowledge and then a novel family of attacks which we call  \emph{forest fire attacks}.

\subsection{Background knowledge}
We assume that attackers know the trustee network in the target service. The reasonableness of this threat model is supported by two evidences. First, attackers can obtain users' usernames. A username is usually a string of letters, digits, and special characters. Moreover, Bonneau et al.~\cite{UserProbing-Bonneau10} showed that a majority (e.g., 96\% in their studies) of websites enable attackers to probe if a string is a legitimate username. Thus, strong attackers, who have enough resources (e.g., a botnet) to perform username probings,  can obtain all usernames in the target service. Second, Schechter et al.~\cite{Schechter09} found, via performing user studies,  that users cannot remember their own trustees. Therefore, a usable trustee-based social authentication system must remind users of their trustees. Recall that an account recovery request only requires a username.  As a result, an attacker could send account recovery requests with the collected usernames to the service provider which reminds the attacker of the trustees of each user. 

Next, we take Facebook as an example to show how an attacker obtains the trustee network. First, Facebook provides an interface\footnote{\url{https://www.facebook.com/login/identify?ctx=recover}} to test if a user (represented by a username, real name, or email address) is in Facebook. Thus, the attacker can perform username probings to collect Facebook users. Second, the attacker sends account recovery requests to Facebook using the collected names. Recall that Facebook shows  all trustees to a user once the user correctly types in one trustee. Moreover, Bilge et al. [4] showed that an attacker can obtain friend lists of around 90\% of Facebook users. Thus, the attacker can repeatedly guess the trustees of a user until success. We note that Facebook only allows a user to try around 10 times for typing in the trustees within a short period of time. However, such rate limit cannot prevent a strong attacker from obtaining the trustee network eventually, though it can increase the attacker's cost.

\subsection{Forest fire attacks}
\label{sec:forest_fire_attack}
Our forest fire attacks consist of  \emph{Ignition Phase} and \emph{Propagation Phase}. 

\myparatight{Ignition Phase} In this phase, an attacker obtains a small number of compromised users which we call  \emph{seed users}. They could be obtained from phishing attacks, statistical guessings, and password database leaks, or they could be a coalition of users who collude each other.  Indeed, a large number of social network accounts were reported to be  compromised,\footnote{For instance, Gao et al.~\cite{Gao10} showed that around 57,000 out of 3,500,000 (1.6\%) Facebook accounts were compromised.}  showing the feasibility of obtaining compromised seed users. 


\myparatight{Propagation Phase} Given the seed users, the attacker iteratively attacks other users. In each \emph{attack iteration}, the attacker performs one \emph{attack trial} to each of the uncompromised users  according to some \emph{attack ordering} of them. In an attack trial to a user $u$, the attacker sends an account recovery request with $u$'s username to the service provider, which issues different verification codes to $u$'s trustees. The goal of the attacker is to obtain verification codes from at least $k$ trustees. If at least $k$ trustees of $u$ are already compromised, the attacker can easily compromise $u$; otherwise, the attacker can impersonate $u$ and send a spoofing message to each uncompromised trustee of $u$ to request the verification code. Schechter et al.~\cite{Schechter09} found that such spoofing attacks can successfully retrieve a verification code with an average probability around 0.05. 


\begin{table*}[!t]\renewcommand{\arraystretch}{1}
\centering
\vspace{-4mm}
\caption{Representative notations used in our model.}
\begin{tabular}{|l|l|} \hline 
{\small  Notations} & {\small Definitions} \\ \hline
{\small $G=(V,E)$} & {\small The trust social network among the users in the service} \\ \hline
{\small $G_T=(V_T,E_T)$} & {\small The trustee network among the users in the service } \\ \hline
{\small $V_a$} &{\small  The set of users who adopt the trustee-based social authentication service} \\ \hline
{\small $u$} &{\small  A user in the service} \\ \hline
{\small $\Gamma(u)$} &{\small  The set of friends of $u$} \\ \hline
{\small $\Gamma_T(u)$} &{\small  The set of trustees of $u$} \\ \hline
{\small $\Gamma_{T,o}(u)$} &{\small  The set of users who select $u$ as a trustee} \\ \hline
{\small $d_{o}(u)$} &{\small  The number  of users who select $u$ as a trustee} \\ \hline
{\small $m_u$} &{\small  The number of trustees of  $u$} \\ \hline
{\small $k$} &{\small  Recovery threshold, i.e., $u$ is authenticated if $u$ obtains verification codes from $k$ of $m_u$ trustees}  \\ \hline
{\small $n_s$} & {\small The number of seed users compromised in the Ignition Phase} \\ \hline
{\small $S$} & {\small The set of seed users that are compromised in the Ignition Phase} \\ \hline
{\small $\mathcal S$} &{\small  The strategy to select the seed users}\\ \hline
{\small $n$} & {\small The number of attack iterations in the Propagation Phase} \\ \hline
{\small $O^{(t)}$} & {\small The  attack ordering according to which the attacker performs attack trials to the users in the $t$th attack iteration} \\ \hline
{\small $\mathcal O$} & {\small The ordering construction strategy} \\ \hline
{\small $p_s^{(t)}(v,u)$} & {\small The probability of obtaining a verification code from $u$'s trustee $v$ via spoofing attacks in the $t$th attack iteration} \\ \hline
{\small $p_s$} & {\small Average spoofing probability} \\ \hline
{\small $p_c^{(t)}(u)$} &{\small  The probability that $u$ is compromised in the $t$th attack iteration}  \\ \hline
{\small $p_c^{(t)}(V_T)$} & {\small The vector of compromise probabilities of all users in the  $t$th attack iteration} \\ \hline
{\small $p_a^{(t)}(u)$} & {\small The probability that $u$ is compromised in at least one attack iteration after $t$ attack iterations} \\ \hline
{\small $p_a^{(t)}(V_T)$} &{\small The vector of aggregate compromise probabilities of all users after  $t$ attack iterations} \\ \hline
{\small $n_c(G_T, k, n_s, n, {\mathcal S}, {\mathcal O})$} & {\small The expected number of compromised users} \\ \hline
{\small $c_I$} & {\small The cost of obtaining the set of compromised seed users in the Ignition Phase} \\ \hline
{\small $c^{(t)}(u)$} & {\small The expected number of spoofing messages that are sent in the attack trial to $u$ in the $t$th attack iteration} \\ \hline
{\small $c_e$} & {\small The average cost per spoofing  message} \\ \hline
{\small $c(G_T, k, n_s, n, {\mathcal S}, {\mathcal O})$} & {\small The expected cost} \\ \hline
{\small $p_r^{(t)}(u)$} & {\small The recovery probability of $u$ in the $t$th attack iteration} \\ \hline
{\small $p_r$} & {\small Average recovery probability} \\ \hline
\end{tabular}
\label{notation}
\vspace{-2mm}
\end{table*}

Although the spoofing attacks can help attackers compromise more users, we want to stress that they are \emph{optional}. We will show in our experiments that an attacker can still compromise a large number of users even if he does not use spoofing attacks to retrieve verification codes in some cases.

\myparatight{Example} Figure~\ref{forest_fire_attack} shows a forest fire attack to a service with 6 users.  Note that a good attack ordering can increase the probability that users are compromised and decrease the number of required spoofing messages (see our experimental results in Section~\ref{sec:eva}). In our example, if the attacker performs attack trials with an attack ordering of $u_5, u_6, u_4$, the attacker needs to spoof both $u_4$ and $u_6$ to compromise $u_5$, which requires two spoofing messages. However, with the attack ordering of $u_6, u_5, u_4$, the attacker only needs to spoof $u_4$ to compromise $u_5$, which only requires one spoofing message and could succeed with a higher probability. 

\myparatight{Compromised users could be recovered} Users could recover their compromised accounts to be uncompromised after they or the service provider detect suspicious activities of the accounts. For instance, a trustee of $u$ receiving a spoofing message might report to $u$, who then changes his or her password; the phenomenon that a trustee requests lots of verification codes for different users within a short period of time is a possible indicator of forest fire attacks, and the service provider could then notify the users, whose trustees have requested verification codes, to change passwords. Moreover, a recovered account could be compromised again in future attack iterations, e.g., when the trustees of the recovered user are still compromised. The process of being compromised and being recovered could repeat for many attack iterations.

\section{Security Model}
\label{sec:model}
In this section, we introduce our security model to formalize the threats  of forest fire attacks and their costs for attackers. 
 Table~\ref{notation} summarizes our important notations.

\subsection{Formalizing threats}
We use $\Gamma_T(u)$ and $m_u=|\Gamma_T(u)|$ to denote the set of trustees and the number of trustees of $u$, respectively. We denote by $\Gamma_{T,o}(u)$ the set of users who select $u$ as a trustee.   We model a set of seed users (denoted as $S$) is obtained by a \emph{seed users selection strategy}, and we denote it as $\mathcal S$. In the $t$th attack iteration, the attacker performs attack trials to uncompromised users according to an attack ordering $O^{(t)}$. The attack orderings are constructed by an \emph{ordering construction strategy} which is denoted as $\mathcal O$. 

We call the probability that $u$ is compromised in the $t$th attack iteration \emph{compromise probability}, and we denote it as $p_c^{(t)}(u)$. $u$ is eventually compromised if it is compromised in at least one attack iteration. Thus, we denote by $p_a^{(t)}(u)$ the probability that $u$ is compromised after $t$ attack iterations, and $p_a^{(t)}(u)$ is called \emph{aggregate compromise probability}. The compromise probabilities in the $t$th attack iteration depend on the aggregate compromise probabilities after $(t-1)$ attack iterations. Moreover, we use $p_c^{(t)}(V_T)$ and $p_a^{(t)}(V_T)$ to represent the vectors of compromise probabilities and aggregate compromise probabilities of all users in $V_T$, respectively. 

Algorithm~\ref{alg:forest_fire_attack} shows our model of forest fire attacks. 
 Next, we elaborate the iterative computations of compromise probabilities and aggregate compromise probabilities. 

\subsubsection{Ignition Phase} If $u$ is a seed user, then $u$'s initial compromise probability is 1, otherwise we model $u$'s initial compromise probability as 0. Formally,  we have the initial compromise probability of  $u$ as follows:
\begin{align}
p_a^{(0)}(u)=p_c^{(0)}(u)=
\begin{cases}
1 \text{ if } u\in S \\
0 \text{ otherwise}
\end{cases}
\label{pc_ignition_phase}
\end{align}


\subsubsection{Propagation Phase} 
The key component is to update the aggregate compromise probability of $u$ when the aggregate compromise probabilities of $u$'s trustees are given.

\myparatight{Obtaining one verification code}
 We denote by $A$ the event that the attacker obtains a verification code from a trustee $v$ of $u$ and by $p^{(t)}(v,u)$ the probability that $A$ happens in the $t$th attack iteration. Moreover, we denote the event that $v$ is already compromised when the attacker attacks $u$ in the $t$th attack iteration as $B$. Then we can represent $p^{(t)}(v,u)$ as: 
\begin{align}
\label{equ:one_code_event}
p^{(t)}(v,u)&=Pr(A) \nonumber \\
		&=Pr(A|B)Pr(B) + Pr(A|^\lnot B)Pr(^\lnot B) 
\end{align}
, where $^\lnot B$ represents that $B$ does not happen. Next, we model $Pr(A|B)$, $Pr(B)$, $Pr(A|^\lnot B)$, and $Pr(^\lnot B)$, respectively.

When $B$ happens, the attacker can obtain a verification code from $v$ with a probability 1, i.e., $Pr(A|B)=1$.  $Pr(B)$ depends on whether the attacker attacks $v$ before $u$ or not. If $v$ is attacked before $u$, then the probability that $B$ happens is $p_a^{(t)}(v)$, otherwise it is $p_a^{(t-1)}(v)$. Formally, we have:
\begin{align}
\label{equ:one_code_B}
Pr(B)=
\begin{cases}
p_a^{(t)}(v) &\text{ if }  v \text{ is ordered before } u \\
p_a^{(t-1)}(v) &\text{ otherwise}
\end{cases}
\end{align}

When $B$ does not happen,  the attacker can impersonate $u$ and send a spoofing message to $v$ to request a verification code.  We call the probability that such spoofing attacks succeed \emph{spoofing probability}.  Spoofing probability might be different for different trustees. A trustee might behave differently to spoofing messages impersonating different users because he or she might have different levels of trusts with the users that are impersonated. Moreover, spoofing probability might be different in different attack iterations because trustees might gradually become aware of the spoofing attacks. Thus, we model the spoofing probability that the attacker obtains a verification code from $v$ in an attack trial to $u$ in the $t$th attack iteration as $p_s^{(t)}(v,u)$, i.e., $Pr(A|^\lnot B)=p_s^{(t)}(v,u)$.

In summary, we have: 
\begin{align}
&p^{(t)}(v,u)= \nonumber \\
&\begin{cases}
p_a^{(t)}(v) + p_s^{(t)}(v,u)(1-p_a^{(t)}(v)) \text{ if } v \text{ is ordered before } u \\
p_a^{(t-1)}(v) + p_s^{(t)}(v,u)(1-p_a^{(t-1)}(v)) \text{ otherwise}
\end{cases} \nonumber
\end{align}


\begin{algorithm}[!t] 
\DontPrintSemicolon 
\KwIn{$G_T$, $k$, $p_s^{(t)}(v,u)$, $n_s$, $n$, ${\mathcal S}$, ${\mathcal O}$,  $c_e$,  $c_I$, and $p_r^{(t)}(u)$.} 
\KwOut{$n_c(G_T, k, n_s, n, {\mathcal S}, {\mathcal O})$, $c(G_T, k, n_s, n, {\mathcal S}, {\mathcal O})$.} 
\Begin{
//Selecting seed users in the Ignition Phase. \;
$S\longleftarrow {\mathcal S}(G_T, n_s)$\;  

//Calculating the compromise probabilities. \;
//Ignition Phase. \;
\For{$u\in V_T$}{
\eIf{$u\in S$}{
	$p_c^{(0)}(u)\longleftarrow 1$ \;	
}{
	$p_c^{(0)}(u)\longleftarrow 0$ \;	
}
$p_a^{(0)}(u) \longleftarrow p_c^{(0)}(u)$ \;
} 
//Propagation Phase.\;
$t \longleftarrow 1$\;
$C \longleftarrow 0$\;
\While{$t \leq n$}{
//Constructing an attack ordering.\;
$O^{(t)} \longleftarrow {\mathcal O}(G_T,p_a^{(t-1)}(V_T)) $\;
\For{$i=0$ to $O^{(t)}.size()-1$}{
       $u \longleftarrow O^{(t)}[i]$\;
	Apply Equation~\ref{local_update_rule} to $u$.
$p_a^{(t)}(u) \longleftarrow 1- (1-p_a^{(t-1)}(u)) (1-p_c^{(t)}(u))$ \;
$p_a^{(t)}(u) \longleftarrow (1-p_r^{(t)}(u)) p_a^{(t)}(u)$ \;

$c^{(t)}(u) \longleftarrow$ Apply  Equation~\ref{equ:one_user_cost}\;
$C \longleftarrow C + c^{(t)}(u)$\;
}
$t \longleftarrow t+1$\;
}

//The expected number of compromised users. \;
$n_c(G_T, k, n_s, n, {\mathcal S}, {\mathcal O}) \longleftarrow \sum_{u\in V_T} p_a^{(n)}(u)$ \;

//The expected cost. \;
$c(G_T, k, n_s, n, {\mathcal S}, {\mathcal O}) \longleftarrow c_I + c_eC$ \;

\Return $n_c(G_T, k, n_s, n, {\mathcal S}, {\mathcal O})$, $c(G_T, k, n_s, n, {\mathcal S}, {\mathcal O})$\;
} 
\caption{Our Model of Forest Fire Attacks} 
\label{alg:forest_fire_attack}
\end{algorithm}

\myparatight{Computing compromise probabilities} Recall that $u$ is compromised if the attacker can obtain verification codes from at least $k$ trustees of $u$. Thus, given the natural assumption that $u$'s trustees are independent,  the compromise probability of $u$ in the $t$th attack iteration is calculated using the following local update rule:
\begin{align}
p_c^{(t)}(u) = \sum_{\phi} \prod_{v \in \phi} p^{(t)}(v,u) \prod_{v\in \Gamma_T(u) -\phi} (1-p^{(t)}(v,u))
\label{local_update_rule}
\end{align}
, where $\phi \subseteq \Gamma_T(u)$ and $|\phi| \geq k$.


\myparatight{Aggregating compromise probabilities} Assuming that whether $u$ is compromised in one attack iteration is independent with whether $u$ is compromised in another attack iteration, we can iteratively compute the aggregate compromise probability of $u$ as follows: 
%
\begin{align}
\label{equ:iterative}
p_a^{(t)}(u)=1-(1-p_a^{(t-1)}(u)) (1-p_c^{(t)}(u))
\end{align}

\myparatight{Compromised users can recover to be uncompromised} As we have discussed in Section~\ref{sec:forest_fire_attack}, compromised users can recover to be uncompromised and be compromised again due to various factors.  We call the probability that a compromised user $u$ recovers to be uncompromised in the $t$th attack iteration  \emph{recovery probability}, and we denote it as $p_r^{(t)}(u)$. Considering the recovery probability, we reformulate the aggregate compromise probability of $u$ (i.e., Equation~\ref{equ:iterative})  as follows:
\begin{align}
p_a^{(t)}(u)=(1-p_r^{(t)}(u))(1-(1-p_a^{(t-1)}(u)) (1-p_c^{(t)}(u))) \nonumber
\end{align}


\myparatight{Expected number of compromised users} Given a forest fire attack specified by the seed users selection strategy $\mathcal S$, the number of seed users $n_s$, the ordering construction strategy $\mathcal O$, and the number of attack iteratons $n$, we define the threat of the attack as the expected number of users it compromises. Moreover, we denote the expected number of compromised users as $n_c(G_T, k, n_s, n, {\mathcal S}, {\mathcal O})$, and it is formalized as follows:
\begin{align}
\label{comp_node}
n_c(G_T, k, n_s, n, {\mathcal S}, {\mathcal O}) = \sum_{u \in V_T} p_a^{(n)}(u)
\end{align}

\subsection{Formalizing costs}
We consider the attacker's costs of obtaining seed users and sending spoofing messages. Suppose the cost to obtain seed users is $c_I$. In the following, we quantify the cost of sending spoofing messages. 

In an attack trial to $u$, the attacker sends a spoofing message to  $u$'s  trustee $v$ when the following three independent events happen: $u$ is uncompromised, $v$ is uncompromised, and among the rest of $u$'s trustees, the number of compromised ones is less than $k$. 

In the $t$th attack iteration, the first event happens with a probability $(1-p_a^{(t-1)}(u))$. We denote by $q^{(t)}(v,u)$ the probability that the second event happens, and $q^{(t)}(v,u)$ depends on  whether $v$ is attacked before $u$  or not. Formally, we have:
 \begin{align}
\label{equ:prob}
&q^{(t)}(v,u)=
&\begin{cases}
1-p_a^{(t)}(v) \text{ if } v \text{ is ordered before } u\\
1-p_a^{(t-1)}(v) \text{ otherwise}
\end{cases}
\end{align}
, for any $v\in \Gamma_T(u)$. Moreover, the third event happens with the following probability:
 \begin{align}
\label{equ:less_than_k}
&r^{(t)}(v,u) = \nonumber \\
&\sum_{\phi} \prod_{w\in \phi} (1-q^{(t)}(w,u)) \prod_{w\in \Gamma_T(u) - \phi - \{v\}} q^{(t)}(w,u)
\end{align}
, where $\phi\subseteq \Gamma_T(u) -\{v\}$ and $|\phi| < k$.

Therefore, the \emph{expected} number of spoofing messages (denoted as $c^{(t)}(v, u)$) that the attacker sends to  $v$ in the attack trial to $u$ in the $t$th attack iteration is formalized as follows:
 \begin{align}
\label{equ:one_cost}
c^{(t)}(v,u)= (1-p_a^{(t-1)}(u))q^{(t)}(v,u)r^{(t)}(v,u)
\end{align}

So the total expected number of spoofing messages (denoted as $c^{(t)}(u)$) that the attacker sends in the attack trial to $u$ in the $t$th attack iteration is:
\begin{align}
\label{equ:one_user_cost}
c^{(t)}(u)= (1-p_a^{(t-1)}(u))\sum_{v\in \Gamma_T(u)} q^{(t)}(v,u)r^{(t)}(v,u)
\end{align}

Thus, we obtain the total expected cost of performing a forest fire attack as follows:
 \begin{align}
\label{equ:total_cost}
c(G_T, k, n_s, n, {\mathcal S}, {\mathcal O})= c_I + c_e\sum_{t=1}^{n} \sum_{u\in V_T}c^{(t)}(u)
\end{align}
, where $c_e$ is the average cost  for the attacker to send one spoofing message. 


\section{Attack Strategies}
\label{sec:attack_strategy}
The attacker could design the seed users selection strategy and the attack ordering construction strategy to maximize the expected number of compromised users. First, we show that finding the optimal set of seed users and the optimal ordering construction strategy  is NP-Complete. Then, we explore various scenarios where seed users have different properties and introduce two ordering construction strategies.

\subsection{NP-Completeness}

Given $G_T,k,n_s,$ and $n$, 
 the attacker essentially aims to solve the following  \emph{attack maximization problem}:
\begin{align}
\label{attack_max}
&\text{\bf Attack Maximization Problem:} \nonumber \\
&n_c(G_T,k,n_s,n)=\max_{{\mathcal S}, {\mathcal O}}n_c(G_T,k,n_s,n,{\mathcal S}, {\mathcal O}) \nonumber
\end{align}

We prove that this problem is NP-Complete. Please refer to Appendix~\ref{theorem:attack:proof} for the proof.


\subsection{Strategies for selecting seed users}
\label{sec:seed_selection}


A seed users selection strategy ${\mathcal S}$ is essentially to assign a score which represents some metric of importance to each user and to select $n_s$ users with the highest scores as seed users. This is closely related to the \emph{node centrality problem}~\cite{nodeCentrality} in the network science community. In the following, we modify a few node centrality heuristics as seed users selection strategies. These strategies work on a trustee network, and we name them with a prefix `S-' to indicate they are used to select seed users. 

\myparatight{S-Random} As a baseline, this strategy assigns a random score ranging from 0 to 1 to each user in the trustee network. 

\myparatight{S-Degree} Intuitively, if a user $u$ is selected as a trustee by more users, then compromising $u$ will increase the compromise probabilities of more users and thus the attacker has an opportunity to compromise more users. Therefore, S-Degree treats the number of users that select $u$ as a trustee (i.e., outdegree of $u$ in the trustee network) as $u$'s score.  

\myparatight{S-BadRank} S-BadRank, adapted from BadRank~\cite{badrank}, performs a random walk on the trustee network. Specifically, S-BadRank starts a random walk from $u$ that is picked uniformly at random. Then S-BadRank iteratively performs one of the two operations: choosing a trustee $v$ of $u$ uniformly at random and walks to $v$ with a probability $1-\alpha$, and selecting a user $w$ uniformly at random from the entire trustee network and walks to $w$ with a probability $\alpha$. Traditionally, $\alpha$ is called the \emph{restart probability}.

The random walk will converge to a stationary probability distribution over all users in the trustee network. The stationary probability of $u$ is roughly the frequency with which the random walk visits $u$ and is used as $u$'s score. Intuitively, S-BadRank tends to assign a high score to a user that is selected by many users as a trustee because he or she has a large chance to be visited by the random walk.  


\myparatight{S-Closeness}  The attacker could also select users that are close to all other users in the trustee network as  seed users because compromising them could help the attacker compromise other users more quickly. This intuition is formally captured by the \emph{closeness} metric~\cite{Sabidussi66}. Specifically, we define the distance between $v$ and $u$ as the length of the shortest directed path whose tail is $v$ and whose head is $u$. Then the closeness of $u$ is defined as the inverse of the sum of its distances to all other users, and it is treated as $u$'s score. In our experiments, we adopt the approximate algorithm developed by Okamoto et al.~\cite{Okamoto08} to find the top-$n_s$ users with the highest closeness scores since it is scalable to large networks.


\subsection{Strategies for constructing attack orderings}
We name these strategies with a prefix `O-' to indicate that they are used to construct attack orderings. Essentially, these strategies assign a score to each user in the trustee network and rank them in a decreasing order according to their scores.

\myparatight{O-Random} As a baseline, this strategy assigns a random score ranging from 0 to 1 to each user in the trustee network.

\begin{algorithm}[!t] 
\DontPrintSemicolon 
\KwIn{$G_T=(V_T, E_T)$ and $p_a^{(t)}(V_T)$.} 
\KwOut{An ordering $O$ of all users  in $V_T$.} 
\Begin{
\For{$u\in V_T$}{
	\For{$v\in \Gamma_T(u)$}{
		//Probability of getting one verification code
		$p_v \longleftarrow p_a^{(t)}(v) + p_s^{(t+1)}(v,u)(1-p_a^{(t)}(v)) $ \;
		
	}
	$p_u\longleftarrow \sum_{\phi} \prod_{v \in \phi} p_v \prod_{v\in \Gamma_T(u) -\phi} (1-p_v)$,  where $\phi \subseteq \Gamma_T(u)$ and $|\phi| \geq k$. \;
	$q_a^{(t)}(u) \longleftarrow 1-(1-p_a^{(t)}(u))(1-p_u)$ \;	
} 
\For{$u\in V_T$}{
	$d_a^{(t)}(u) \longleftarrow q_a^{(t)}(u) - p_a^{(t)}(u)$ \;
} 
$O \longleftarrow$ Sort users decreasingly using $d_a^{(t)}(u)$\;
\Return $O$\;
} 
\caption{O-Gradient} 
\label{alg:gradient}
\end{algorithm}

\myparatight{O-Gradient} Algorithm~\ref{alg:gradient} shows O-Gradient. Given the current aggregate compromise probabilities $p_a^{(t)}(V_T)$ of all users after $t$ attack iterations, the attacker calculates a \emph{predicted} aggregate compromise probability $q_a^{(t)}(u)$ for each user $u$ by simulating an attack trial to $u$. Note that $q_a^{(t)}(u)$ might be different from $p_a^{(t+1)}(u)$ because $q_a^{(t)}(u)$ is calculated with the fixed aggregate compromise probabilities of $u$'s trustees while $p_a^{(t+1)}(u)$ is updated with the newest aggregate compromise probabilities of $u$'s trustees.     
Intuitively, a user $u$ with a larger difference of  $q_a^{(t)}(u)-p_a^{(t)}(u)$ could be attacked with a higher priority because such an attack trial could bring more increase to the aggregate compromise probabilities of users who select $u$ as a trustee.  Thus, O-Gradient calculates the differences of $q_a^{(t)}(u)-p_a^{(t)}(u)$ for all users and ranks them decreasingly according to these differences.     


\section{Defense Strategies}
\label{sec:defense_strategy}
We discuss potential defense strategies from three aspects, i.e., hiding trustee networks from attackers, mitigating spoofing attacks, and constraining the selection of trustees.   


\subsection{Hiding trustee networks}
Preventing attackers from obtaining a trustee network is an essential step towards the defense of forest fire attacks. In the currently deployed social authentication systems, attackers can obtain a trustee network because \emph{users need to know their trustees to retrieve verification codes from them} in the Recovery Phase.   An alternative implementation of the Recovery Phase is that the service provider directly sends verification codes to the trustees of the user when receiving an account recovery request, and the trustees are required to actively share the verification codes to the user.  This implementation does not require users to know their trustees, and thus it is hard for attackers to obtain the trustee network. 


However, this implementation  is \emph{unreliable} and could annoy users and their trustees. Specifically,  $u$'s trustees might already forget they are trustees of $u$, and thus they might simply treat those verification codes as spams and not share them with $u$, which results in low reliability. Moreover, users do forget who their trustees are~\cite{Schechter09}, and thus it is highly impossible for $u$ to actively request verification codes from its trustees. If the trustees do actively share the codes with $u$, then attackers can frequently send account recovery requests to the service provider which immediately sends verification codes to the trustees, and the trustees will (possibly) frequently share the codes with $u$,  which could be annoying  to both $u$ and  its trustees. More seriously,  after $u$'s trustees realize that the verification codes are just spams, they might not actively share the verification codes with $u$ even if she is really trying to recover her account, which again results in low reliability.

Therefore, hiding trustee networks from attackers sacrifices reliability, which possibly explains why existing trustee-based social authentication systems didn't adopt this alternative implementation of the Recovery Phase. 

  
\subsection{Mitigating spoofing attacks}
Another way to defend against forest fire attacks is to remind trustees of not sharing verification codes via messages. This strategy is not novel, and we include it for completeness. Indeed, existing social authentication systems~\cite{Schechter09,facebookSocAuth} already try to mitigate spoofing attacks.  For instance,  Microsoft's system~\cite{Schechter09} asks a trustee why she is requesting the verification code and encourages her to share the code with the user via phone or meeting in person. 

However, it is hard to control how trustees share the verification codes with others in practice. Indeed,  an attacker can still obtain a verification code from a trustee with an average probability around 0.05 via message-based spoofing attacks in the Microsoft's system~\cite{Schechter09}. It is an interesting future work to design better user interfaces to further reduce this spoofing probability.

\subsection{Constraining trustee selections}
Finally, we introduce strategies to contrain trustee selections, which are easy to implement and effective at defending against forest fire attacks. We consider both \emph{local trustee selection strategies}  and \emph{global trustee selection strategies}. A local trustee selection strategy is based on a user's local social network structure while a global one is based on the entire social network structure.  We name these strategies with a prefix `T-' to indicate that they are used to select trustees. 

We note that how users select their trustees in a real trustee-based social authentication system such as Facebook's Trusteed Contacts is not clear and thus might not be one of our strategies. However, our work focuses on a comparative study about different trustee selection strategies and can shed light on which strategy is more secure. 


\subsubsection{Local trustee selection strategies} For a user $u$, a local trustee selection strategy essentially computes a score $s(v,u)$ for each friend $v$ of $u$ and then selects $m_u$ friends with the highest scores as $u$'s trustees.  

\myparatight{T-Random} As a baseline, T-Random assigns a random number ranging from 0 to 1 as the score $s(v,u)$. 
 
\myparatight{T-CF} As was shown by Gilbert and Karahalios~\cite{gilbert:chi09}, the number of common friends of two users is an informative indicator about the level of trust between them. Thus, one speculation is that a user might select friends with which he or she shares many common friends as trustees. To quantify the security of this speculation, we design the strategy T-CF (i.e.,T-CommonFriends), which uses the number of common friends shared by $u$ and his or her friend $v$ as the score $s(v,u)$, i.e., $s(v,u) = |\Gamma(v) \cap \Gamma(u)|$.

However, there are two drawbacks of T-CF. First, the fact that $u$ shares many friends with a popular user $v$ doesn't necessarily mean that $u$ and $v$ have a high level of trust because it is normal for many friends of $u$ to be in $v$'s friend list. 
 Second, if a common friend of $u$ and $v$ is a  popular user, then sharing him or her doesn't necessarily indicate a high level of trust between $u$ and $v$. Next, we introduce two strategies to overcome the two drawbacks, respectively.

\myparatight{T-JC} To overcome the first drawback of T-CF, we design the trustee selection strategy T-JC (i.e., T-JaccardCoefficient), which uses the Jaccard Coefficient~\cite{Jaccard1901} of the two sets $\Gamma(u)$ and $\Gamma(v)$ as the score $s(v,u)$, i.e., $s(v,u) = \frac{|\Gamma(v) \cap \Gamma(u)|}{|\Gamma(v) \cup \Gamma(u)|}$.
 

\myparatight{T-AA} To overcome the second drawback, we design the T-AA (i.e., T-AdamicAda) strategy, which uses Adamic-Ada~\cite{aa03} similarity between $u$ and $v$ as the score  $s(v,u)$. Adamic-Ada similarity penalizes each common friend by its popularity (i.e., the number of friends). Formally, we have
$s(v,u) = \sum_{w\in \Gamma(v) \cap \Gamma(u)} \frac{1}{\text{log}|\Gamma(w)|}$.


\begin{algorithm}[!t] 
\DontPrintSemicolon 
\KwIn{$G=(V, E)$, $V_a$, and $m_u$ of all $u\in V$.} 
\KwOut{A trustee network $G_T=(V_T, E_T)$.} 
\Begin{
$V_T \longleftarrow V$ \;
\For{$u\in V_T$}{
$d_o(u) \longleftarrow 0$ \;
}
\For{$u\in V_a$}{
$\phi \longleftarrow \emptyset$ \;
\While {$|\phi| < m_u$}{
	$v \longleftarrow \argmin_{v\in \Gamma(u)-\phi} d_o(v)$ \;
	$\phi \longleftarrow \phi \cup \{v\}$ \;
	$d_o(v) \longleftarrow d_o(v) + 1$ \;
	$E_T \longleftarrow E_T \cup (v, u)$ \;
}
} 

\Return $G_T=(V_T, E_T)$\;
} 
\caption{T-Degree} 
\label{alg:d-degree}
\end{algorithm}

\subsubsection{Global trustee selection strategies}  Global strategies leverage the entire social network structure and thus are potentially better than local strategies.
 
\myparatight{T-Degree} As we discussed in Section~\ref{sec:seed_selection}, seed users could be those having large outdegrees in the trustee network, and they could enable an attacker to compromise many other users. Thus, we propose the T-Degree  strategy to minimize the maximum outdegree in the trustee network.  Intuitively, T-Degree constrains that no users are selected as trustees by too many other users.

Algorithm~\ref{alg:d-degree} shows our T-Degree strategy. 
T-Degree selects trustees for users one by one. For each user $u$ that has adopted the trustee-based social authentication service, T-Degree selects his or her $m_u$ friends whose current outdegrees in the trustee network are the smallest as $u$'s trustees; ties are broken uniform at random.

%


\section{Experiments}
\label{sec:eva}

\subsection{Experimental setups}

\myparatight{Parameter settings} Unless otherwise mentioned, we set $n=10$, $m_u=m=5$, $k=3$,\footnote{This is the recovery threshold chosen by the Microsoft's and Facebook's trustee-based social authentication systems.} $n_s=1,000$, $p_r^{(t)}(u)=p_r=0$ for every $u$, and $\alpha=0.9$;\footnote{We tried $\alpha$ from 0 to 0.9 with a step size 0.1,  and we found that  S-BadRank with $\alpha=0.9$ achieves the largest expected number of compromised users.} we use the O-Gradient strategy to construct attack orderings, i.e., $\mathcal O$=O-Gradient; according to  Schechter et al.~\cite{Schechter09}, the average message-based spoofing probability is around 0.05, thus we set $p_s^{(t)}(v,u)=p_s=0.05$ for every edge $(v,u)\in E_T$.

\begin{table}[!t]\renewcommand{\arraystretch}{1.3}
\centering
\caption{The number of users, the number of users who have at least 10 friends, the fraction of such users, and the average number of friends of such users in the three social networks.}
\begin{tabular}{|c|c|c|c|} \hline 
 & {\small Flickr} & {\small Google+} & {\small Twitter} \\ \hline
{\small \# Nodes} &{\small 1,551,824} &{\small 10,230,332} &{\small 21,297,771} \\ \hline
{\small \# Nodes (degree $\geq$ 10)} &{\small 233,067} &{\small 3,144,370} &{\small 4,540,483} \\ \hline
{\small Fractions} &{\small 15\%} &{\small 31\%} &{\small 21\%} \\ \hline
{\small Average degrees} &{\small 75}&{\small 27}&{\small 107}\\ \hline
\end{tabular}
\label{dataset}
\end{table}

\begin{figure}[!ht]
\vspace{-2mm}
\centering
\subfloat[Number of compromised users]{\includegraphics[width=0.25\textwidth, height=1.5in]{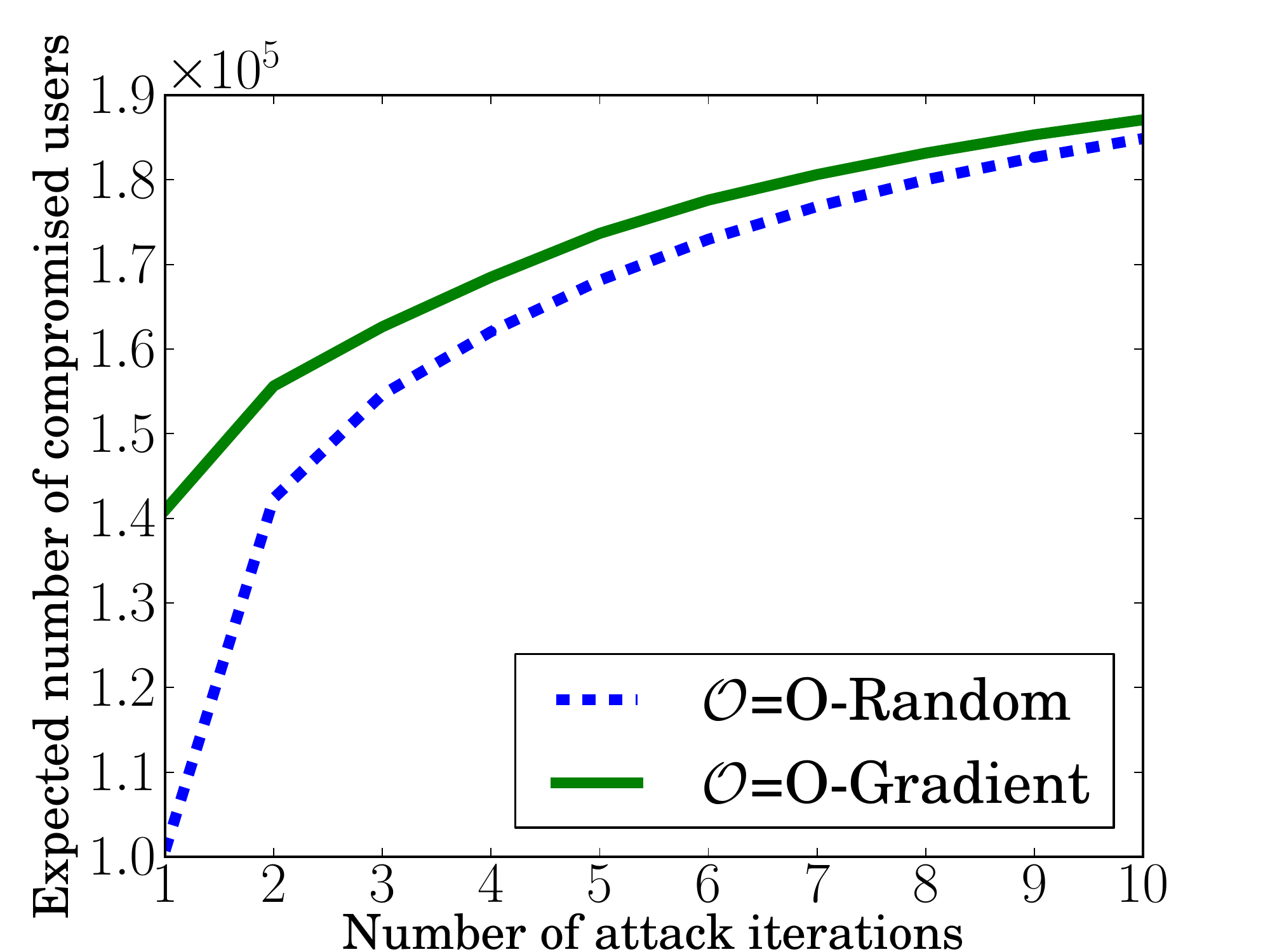}\label{impact-a}}
\subfloat[Number of spoofing messages]{\includegraphics[width=0.25\textwidth, height=1.5in]{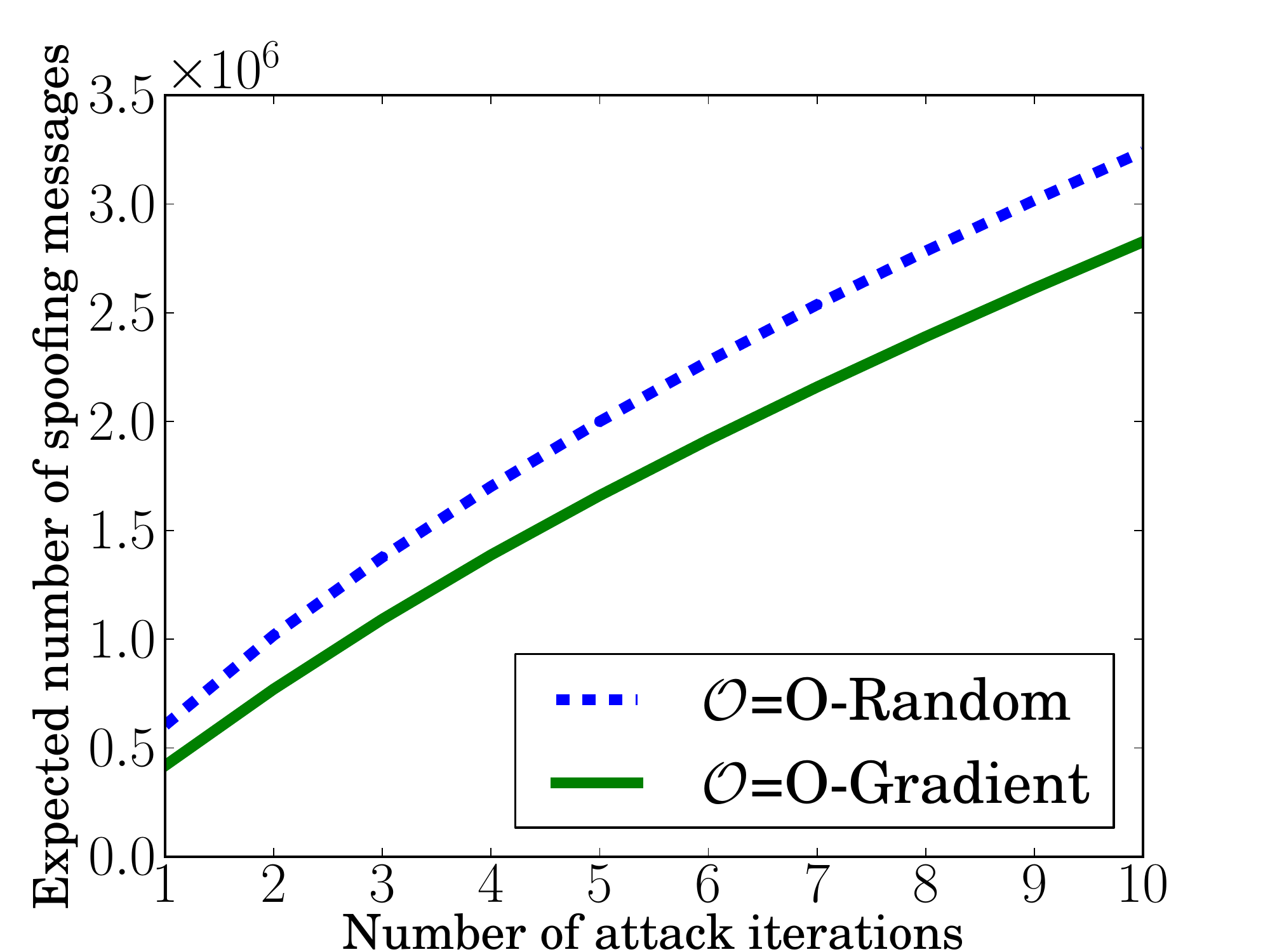}\label{impact-a-cost}}
\caption{Impact of  attackers' resources and  attack orderings on Flickr. First, we find that an attacker can perform forest fire attacks with low costs. Second, we find that O-Gradient compromises more users and requires fewer spoofing messages  than O-Random when the attacker performs a given number of attack iterations.}
\vspace{-4mm}
\label{impact-a-both}
\end{figure}

\begin{figure*}[!ht]
\centering
\vspace{-4mm}
\subfloat[Flickr]{\includegraphics[width=0.33\textwidth, height=1.8in]{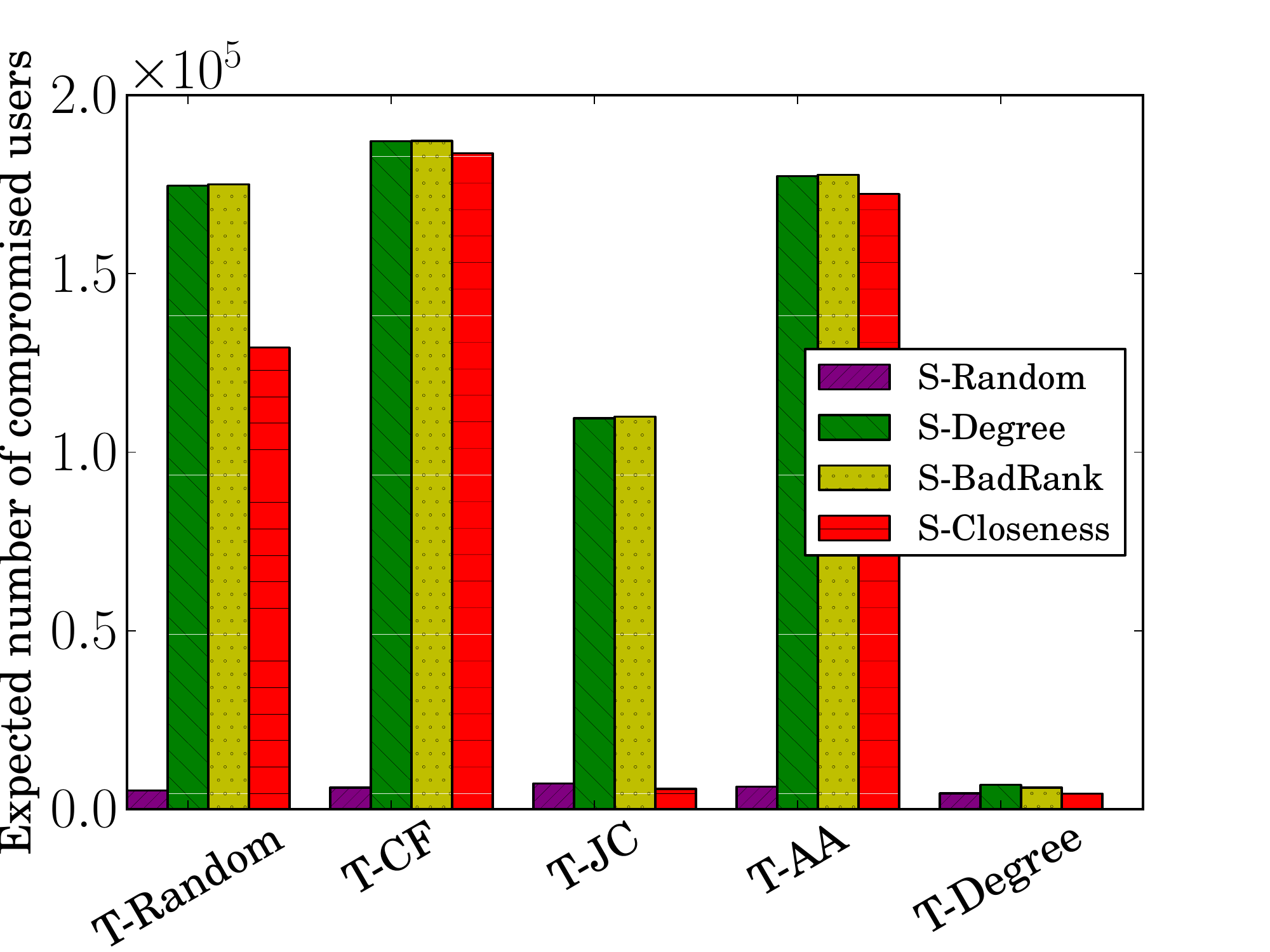}\label{liveJournal-attack-defense}}
\subfloat[Google+]{\includegraphics[width=0.33\textwidth, height=1.8in]{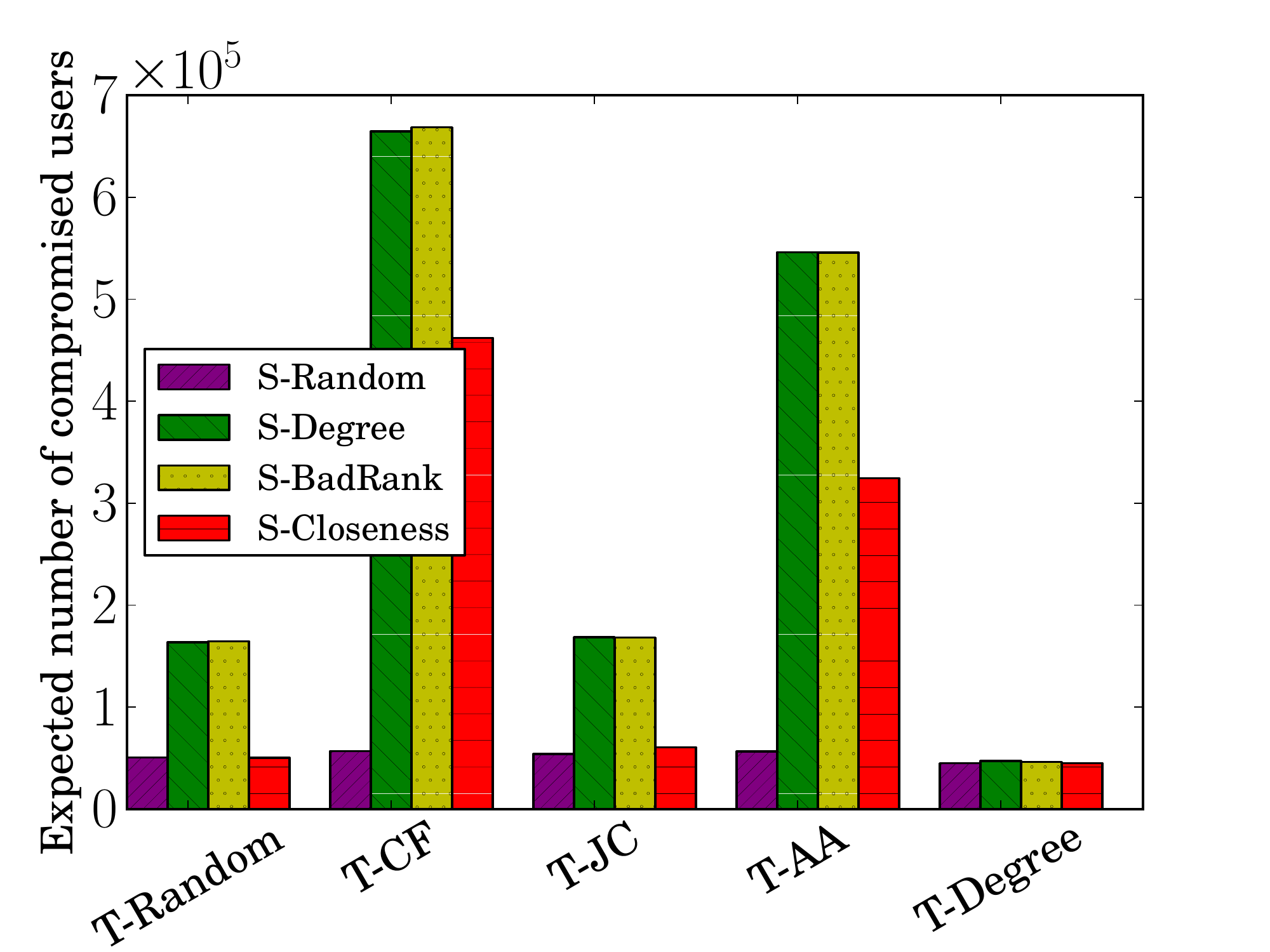}\label{google-attack-defense}}
\subfloat[Twitter]{\includegraphics[width=0.33\textwidth, height=1.8in]{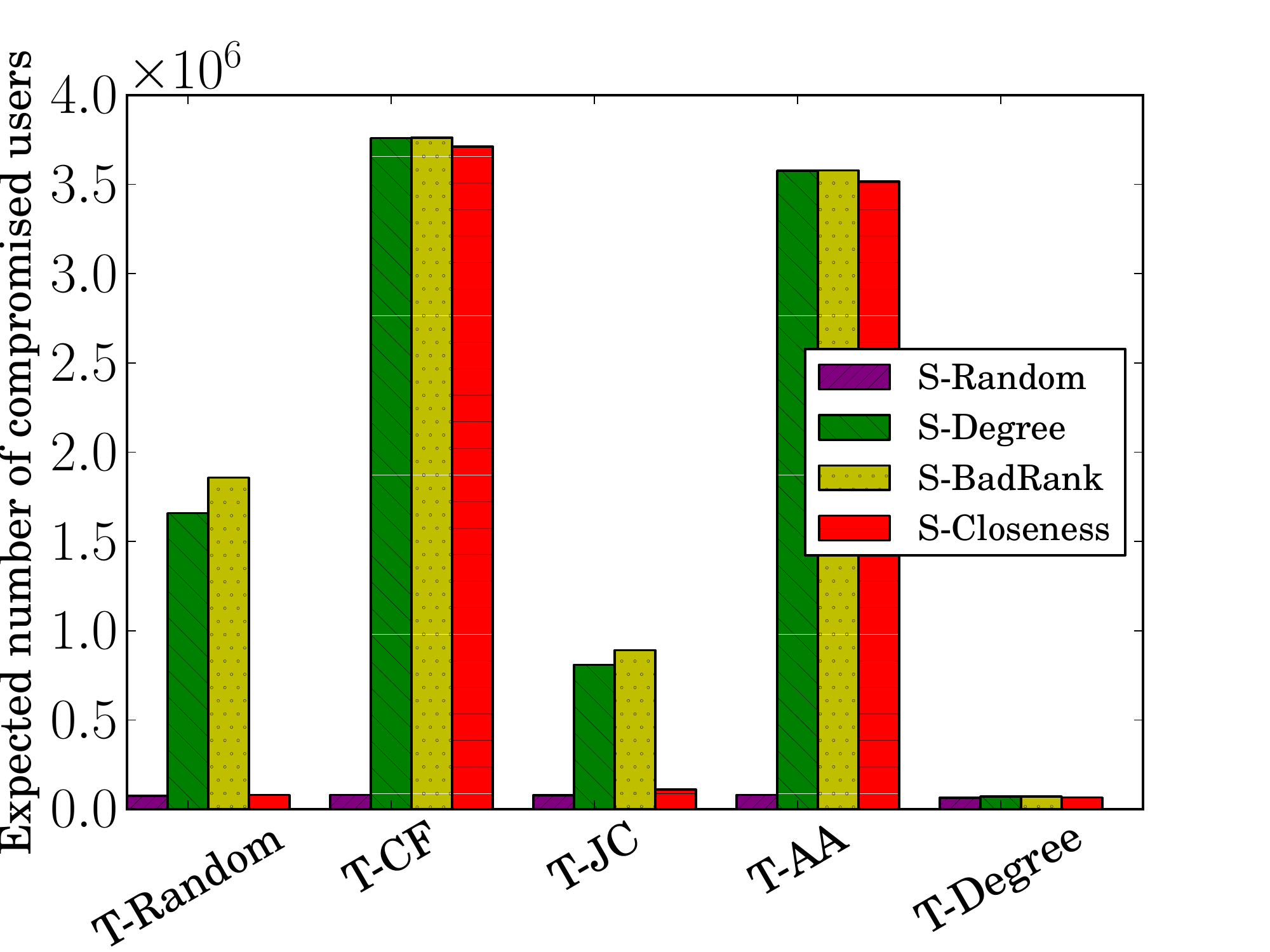}\label{twitter-attack-defense}}
\caption{The expected number of compromised users for different seed users selection strategies and trustee selection strategies. We find that, with 1,000 seed users, an attacker can compromise two to three orders of magnitude more users with low costs of sending spoofing messages in some cases. However, our   strategy T-Degree can decrease the expected number of compromised users by one to two orders of magnitude.}
\label{attack-defense}
\vspace{-4mm}
\end{figure*}

\myparatight{Datasets} We use three social network datasets. They are Flickr, Google+, and Twitter. We obtained the Flickr dataset from Mislove et al.~\cite{Mislove08}. In this social network, we take each user as a node and connect two users if they are in each other's friend lists.  We obtained a Google+ snapshot from Gong et al.~\cite{Gong12-imc}. In this social network, each node is a Google+ user and two users are connected if they are in each other's circles.  The Twitter dataset was obtained from Kwak et al.~\cite{Kwak10}. In this social network,  each node is a Twitter user and two users are connected if they follow each other.

Since a trustee-based social authentication requires a user to have enough number of friends from which trustees can be selected, we assume that users who have at least 10 friends are appropriate to adopt  the trustee-based social authentication service. In our experiments, we consider the \emph{worst} scenario in which every such user has adopted the trustee-based social authentication service.\footnote{The fewer users adopt the trustee-based social authentication, the more secure it is. An extreme case is that if no user adopts the system, then the forest fire attacks cannot be spreaded from the seed users at all.}  Table~\ref{dataset} shows the statistics of our interest in the three social networks.

For simplicity, we only show results on the Flickr dataset in some of our experiments, but similar conclusions are applicable to the other two datasets.



\subsection{Experimental results}
\myparatight{Impact of  attackers' resources and  attack orderings}
The number of attack iterations is closely related to the costs of sending spoofing messages. Figure~\ref{impact-a-both} illustrates the expected number of compromised users (Figure~\ref{impact-a}) and the expected number of required spoofing messages (Figure~\ref{impact-a-cost}) as a function of the number of attack iterations on Flickr for the ordering construction strategies O-Random and O-Gradient. The seed selection strategy is S-Degree and the trustee selection strategy is T-CF.   

First, we find that a strong attacker (e.g.,  an attacker controlling a botnet) can perform firest fire attacks with low costs. This is because such an attacker can send out billions of messages \emph{per day} with a low cost~\cite{emailcost}, which is far more than that needed to spoof trustees in our experiments. 



Second, we find that the ordering construction strategy O-Gradient compromises more users and requires fewer spoofing messages  than O-Random when the attacker performs a given number of attack iterations. 
In other words, given the number of spoofing messages the attacker can send,  the attacker should adopt O-Gradient  to construct the attack orderings. 

Similar conclusions are applicable to other combinations of seed selection strategies and trustee selection strategies. Thus, we don't show the corresponding results for conciseness.



\begin{figure}[!ht]
\centering
{\includegraphics[width=0.4\textwidth, height=1.8in]{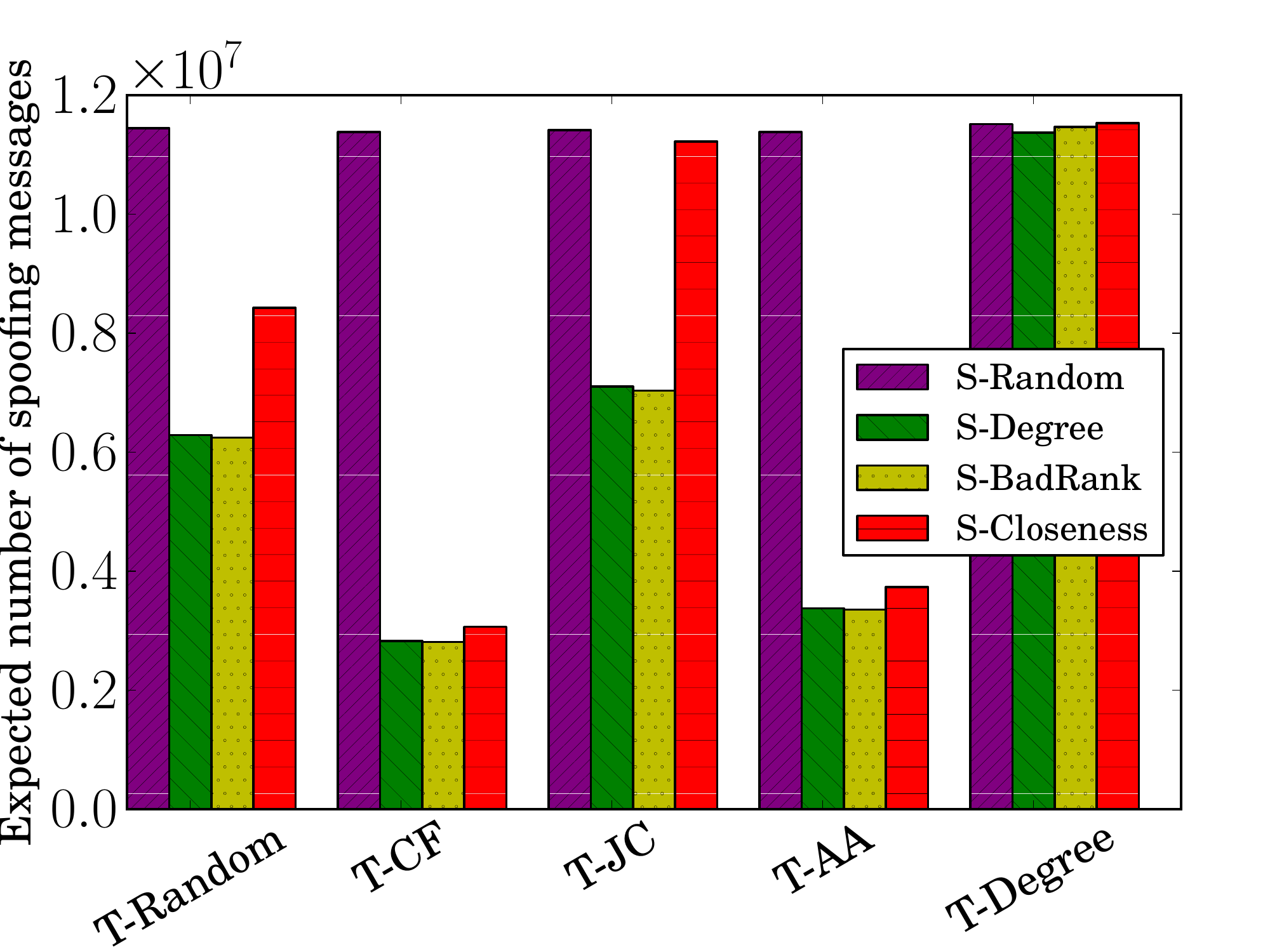}}
\caption{The expected number of required spoofing messages for different seed users selection strategies and trustee selection strategies on Flickr. We observe that our strategy T-Degree can increase the attacker's costs by a few times in some cases.}
\label{attack-defense-cost}
\end{figure}

\myparatight{Impact of seed users selection strategies and trustee selection strategies} 
Figure~\ref{attack-defense} and Figure~\ref{attack-defense-cost} show the expected number of compromised users and the expected number of required spoofing messages respectively for different seed selection strategies and trustee selection strategies. 
We can draw a few conclusions.

First, we find that forest fire attack is a potential big threat. For instance, when the seed users are appointed as trustees of many users (i.e., S-Degree) and the trustees are selected by T-CF, the attacker can compromise around 190,000, 660,000, and 3,760,000 users in the three social networks, 
 respectively. This represents a growth of two to three orders of magnitude from the 1,000 seed users. However, our strategy T-Degree can decrease the expected number of compromised users by one to two orders of magnitude. For instance, the expected number of compromised users of T-Degree is 53 times smaller than that of T-CF on Twitter when the seed users selection strategy is S-Degree. Moreover, our  strategy T-Degree can increase the costs for attackers by a few times in some cases. For instance, the cost of sending spoofing messages of T-Degree is 3 times bigger than that of T-CF and that of T-AA on Flickr when the seed users are appointed as trustees of many users. The reason is that the trustee networks constructed by T-Degree are more loosely connected, which makes it harder for forest fire attacks to propagate among them.


Second, even if the  seed users are distributed among  a social network uniformly at random (i.e., S-Random), the attacker  can still compromise dozens of times more users. For instance, the attacker can still compromise 65 to 80 times more users in Twitter depending on how trustees are selected.   


Third, T-JC works better than T-AA which performs better than T-CF for all seed users selection strategies except S-Random. We find that the outdegree distributions of the trustee networks constructed by T-CF are skewed towards high degrees the most while those constructed by T-JC are skewed towards low degrees the most. Thus, the seed users in the trustee networks constructed by T-JC have lower outdegrees than  those  constructed by T-AA, which have lower outdegrees than those constructed by T-CF. As a result, T-JC performs better than T-AA and T-AA performs better than T-CF.  


\begin{figure}[!t]
\centering
{\includegraphics[width=0.4\textwidth, height=1.8in]{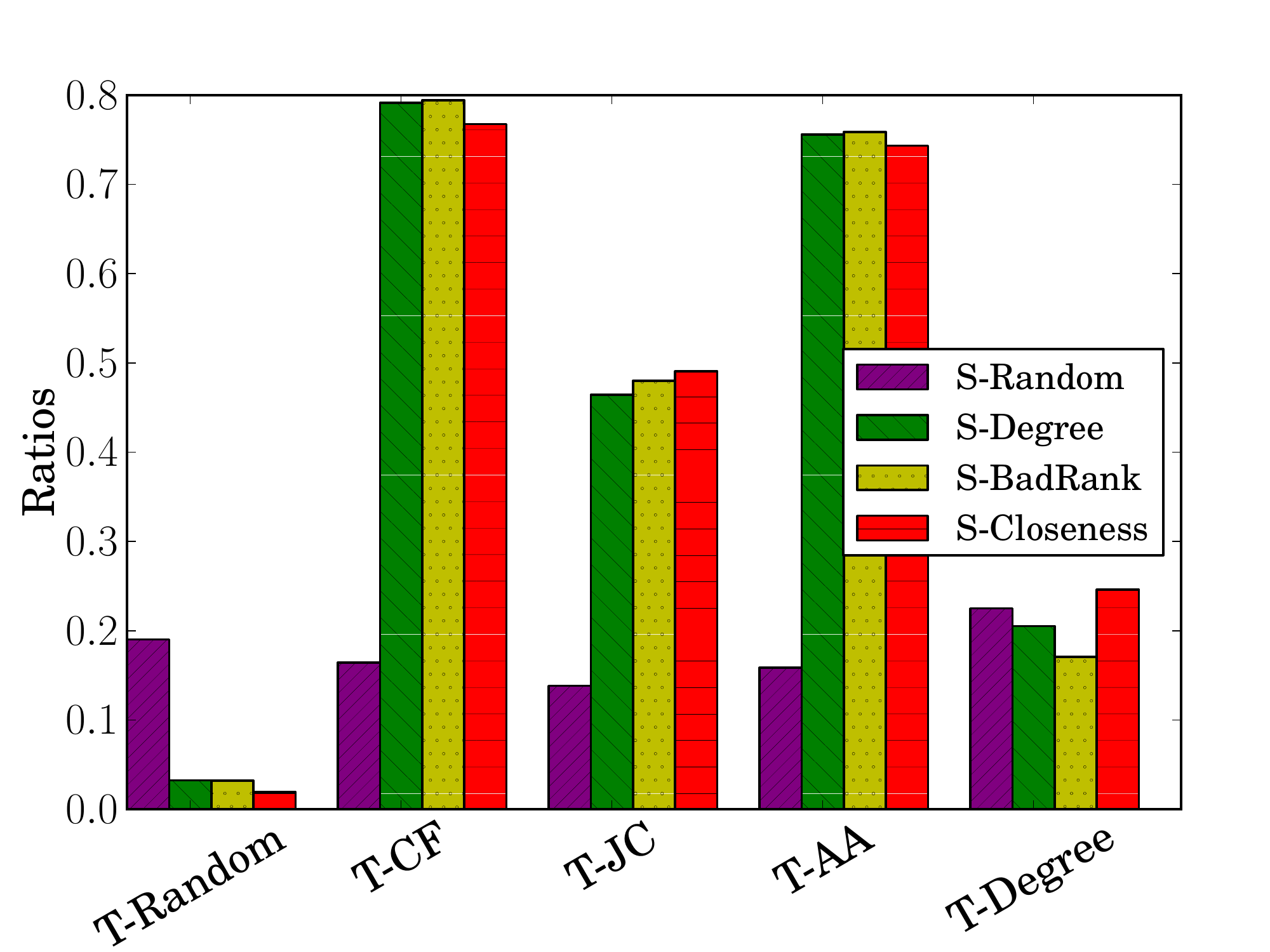}}
\caption{The ratios between the expected number of  users that are compromised without spoofing attacks and those with spoofing attacks for different seed selection strategies and different trustee selection strategies on Flickr. We find that the expected number of compromised users only decreases by 20\% to 25\% when the attacker does not use spoofing attacks in some cases, which implies that spoofing attacks are optional.}
\label{optional-spoofing-attacks}
\end{figure}

\myparatight{Impact of spoofing attacks} The attacker can choose not to send any spoofing message, which requires no costs of sending spoofing messages. Moreover, spoofing probabilities might decrease as the attacker performs more attack iterations because trustees become aware of the spoofing attacks, and we consider an extreme case where spoofing probabilities are zero in all attack iterations.  That spoofing probabilities are zero is equivalent to that the attacker does not use spoofing attacks in terms of the number of compromised users. 

Figure~\ref{optional-spoofing-attacks} shows the ratios between the expected number of  users that are compromised without spoofing attacks and those with spoofing attacks for different seed users selection strategies and  trustee selection strategies on Flickr.

We find that, without spoofing attacks, 
 the expected number of compromised users only decreases by 20\% to 25\% when the trustee selection strategy is T-CF or T-AA and the seed users selection strategy is not S-Random, which means that the attacker can still compromise  orders of magnitude more users and that spoofing attacks are optional in some cases.



\begin{figure}[!t]
\centering
\subfloat[Impact of $m$]{\includegraphics[width=0.25\textwidth, height=1.5in]{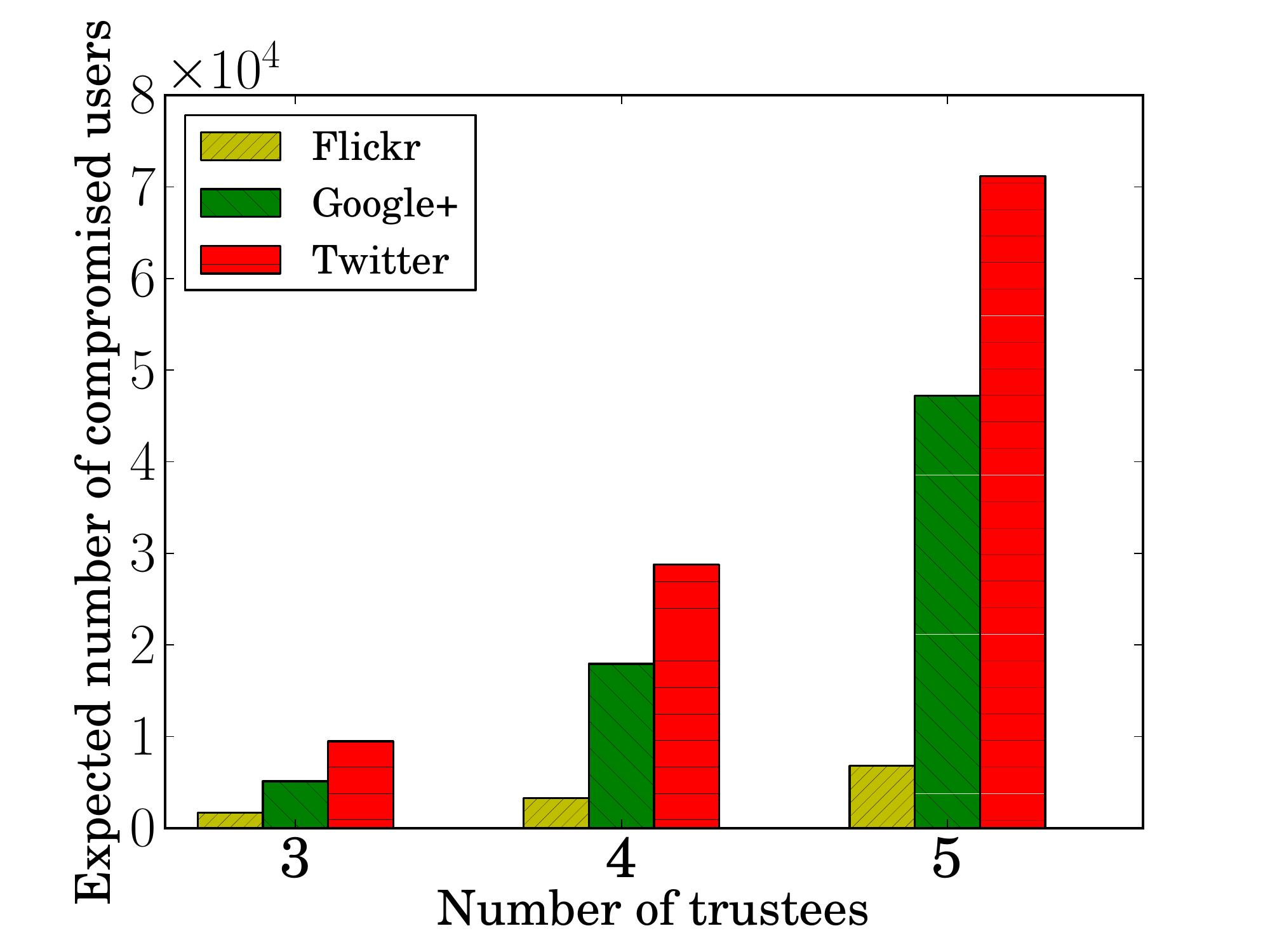}\label{impact-m}}
\subfloat[Impact of $k$]{\includegraphics[width=0.25\textwidth, height=1.5in]{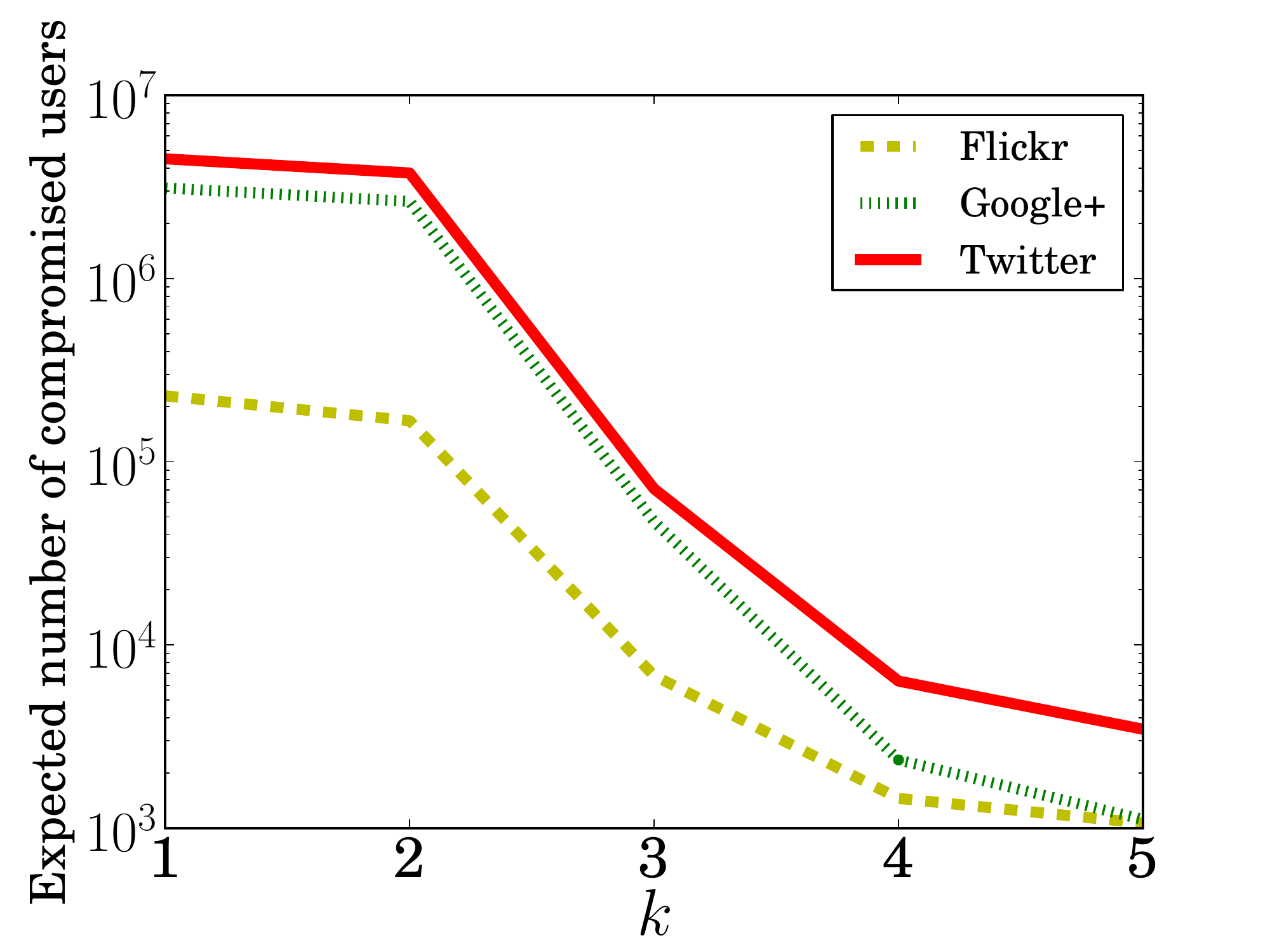}\label{impact-k}}

\subfloat[Impact of $p_s$]{\includegraphics[width=0.25\textwidth, height=1.5in]{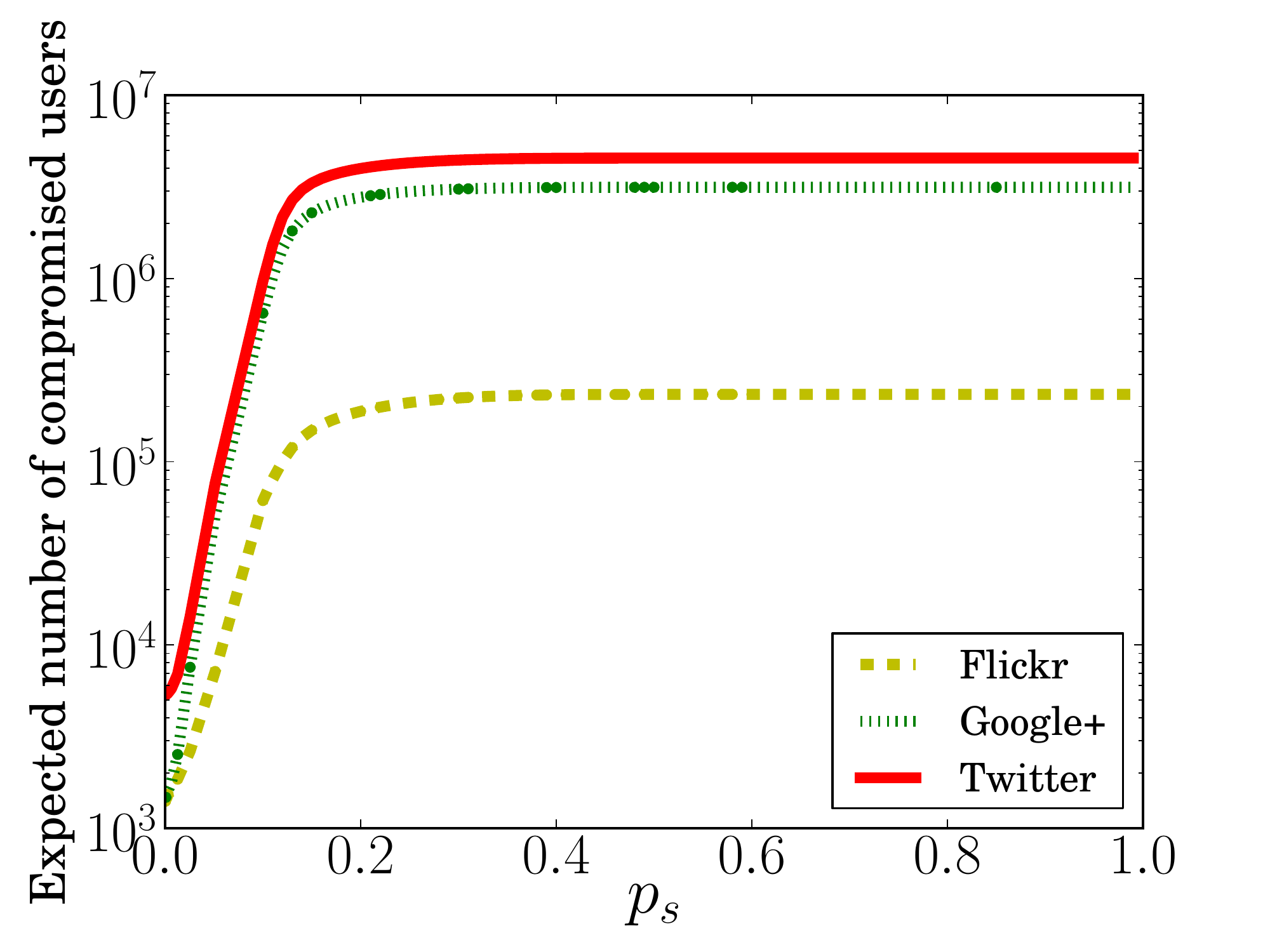}\label{impact-ps}}
\subfloat[Impact of $n_s$]{\includegraphics[width=0.25\textwidth, height=1.5in]{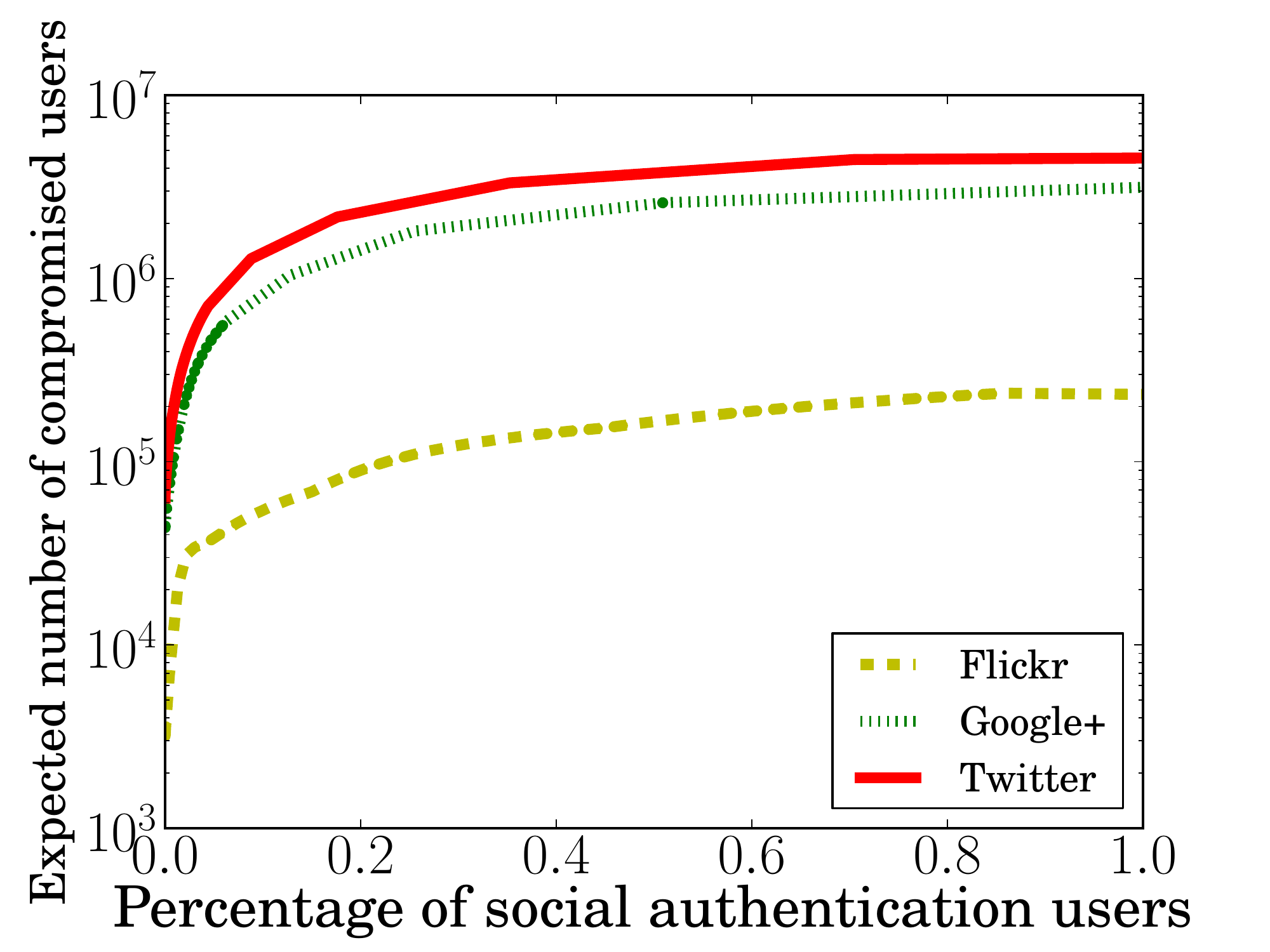}\label{impact-ns}}

\subfloat[Impact of $p_r$]{\includegraphics[width=0.4\textwidth, height=1.8in]{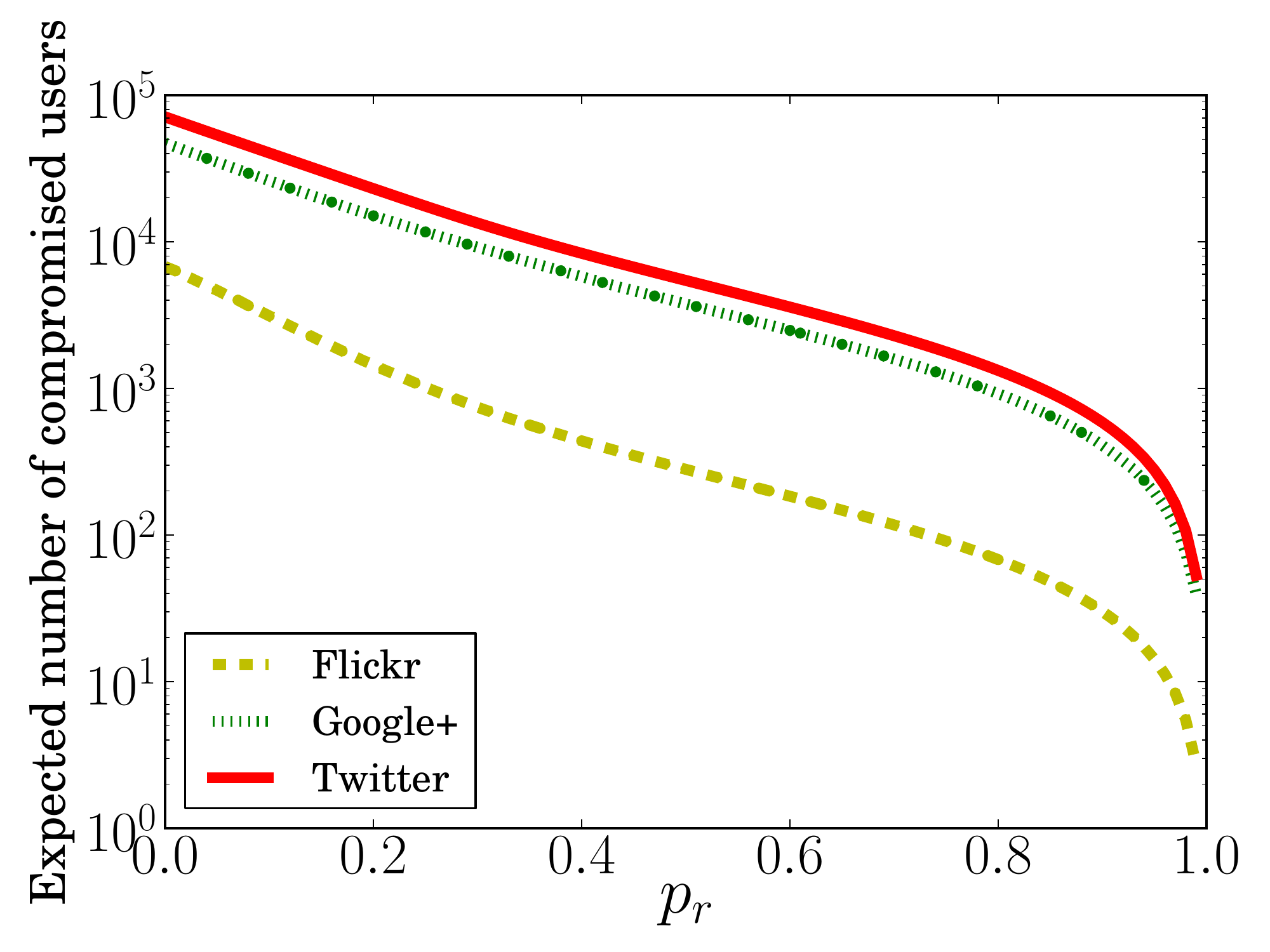}\label{impact-rp}}
\caption{Impact of $m$, $k$, $p_s$,  $n_s$, and $p_r$. 
 We find that $k=4$ is a good tradeoff between security and usability.  
}
\label{phase_transition}
\end{figure}

\myparatight{Impact of {\bf $m_u$, $k$, $p_s^{(t)}(v,u)$,  $n_s$,  and $p_r^{(t)}(u)$}} Towards this end, we assume that the attacker and the defender are making their best choices, i.e., the seed selection strategy is S-Degree and the trustee selection strategy is T-Degree. 
For simplicity, we assume $m_u=m$ for all users $u$ who adopt the trustee-based social authentication service and $p_s^{(t)}(v,u)=p_s$ for all edges $(v,u)$ in the trustee network and attack iterations. Moreover, we consider zeroth-order approximation of recovery probabilities, i.e., we use the average recover probability to approximate recover probabilities of all users in different attack iterations. In other words, $p_r^{(t)}(u)=p_r$ for all users and attack iterations. $p_r$ represents the average fraction of compromised users that realize their accounts are compromised and take actions to recover them in each attack iteration. 

   Figure~\ref{phase_transition} shows the impact of $m$, $k$, $p_s$,  $n_s$, and $p_r$. Note that the x-axis in Figure~\ref{impact-ns} is the percentage of users that adopt the trustee-based social authentication service.  We have  a few observations. 

Naturally, we observe that the expected number of compromised users decreases with $k$ and $p_r$, and it  increases with $m$, $p_s$, and $n_s$. Moreover, the trends are \emph{qualitatively} similar on the three social networks for each of the five parameters, which indicates the generality of our results.


The parameters $m$ and $k$ balance between security and usability. Specifically, a smaller $m$ or a bigger $k$ makes the system more secure but less usable.  Figure~\ref{impact-k} shows that the expected number of compromised users decreases dramatically when $k$ increases from 1 to 4. For instance, the expected number of compromised users decreases by around one order when $k$ changes from 3 to 4. However,  the decrease is not significant when $k$ increases from 4 to 5.  Existing trustee-based social authentication systems~\cite{Schechter09,facebookSocAuth} set $k$ to be 3; our observations indicate that they could set $k$ to be 4 to better balance between security and usability.

We find that the expected number of compromised users changes dramatically when $p_s$ is around 0.2 on all the three social networks. Moreover, almost all users that have adopted the trustee-based social authentication  service are compromised when $p_s$ is bigger than around 0.3.   Our results imply that it is urgent to mitigate the spoofing attacks via designing better user interfaces to remind users of not easily sharing the verification codes to others via messages. 

The expected number of compromised users increases dramatically with $n_s$ in the range where $n_s$ is small. This implies that obtaining a few more seed users enables the attacker to compromise significantly more users when the number of seed users is small. Moreover, even if the number of seed users is 0, the attacker can still compromise thousands of nodes in Flickr and tens of thousands of nodes in Google+ and Twitter.

The expected number of compromised users significantly decreases with $p_r$. For instance, the number of compromised users decreases by one order in all the three social networks when around 40\% of compromised users recover to be uncompromised in each attack iteration (i.e., $p_r=0.4$). Our results imply that it is important for users to take actions to recover their accounts quickly when suspicious activities of their accounts are detected.

\section{Discussion}
\label{sec:discussion}

\myparatight{A greedy seed users selection strategy} An attacker could use a greedy strategy to select seed users. Specifically,  the attacker selects  seed users one by one.  To select a seed user, the attacker iterates over each user that is not a seed user yet; for each such user $u$, the attacker pretends that $u$ is a seed user and  simulates our security model to predict the corresponding expected number of compromised users; and the user $u$ which increases the expected number of compromised users by the most is added as a new seed user. 

We compared this greedy strategy with other seed users selection strategies using a small subnetwork sampled from the Google+ dataset and found that it can compromise more users. However, it is not scalable to large social networks. It is an interesting future work to make the strategy scalable.     

\myparatight{Community-based trustee selection strategy}  We can also design a trustee selection strategy based on some notion of community. Specifically, we could select trustees for users such that the trustee network consists of isolated communities, which could constrain the propagation of forest fire attacks. It is an interesting future work to explore these community-based trustee selection strategies.

\myparatight{Limiting the issuance of verification codes}
A compromised user $u$ might request a verification code from the service provider when the attacker performs an attack trial to a user who selects $u$ as a trustee. Thus, a compromised user who is a trustee of many other users might request many verification codes in each attack iteration.   Therefore, limiting the number of verification codes that each user can request within a given period of time (e.g., one hour)  can slow down forest fire attacks. It is an interesting future work to explore the impact of such rate limitings on  forest fire attacks and what strategies attackers can adopt to maximize the number of compromised users given a time constraint. 


\section{Related Work}
\label{sec:related_work}
\subsection{Social authentications}
Depending on how friends are involved in the authentication process, social authentications can be classified into \emph{trustee-based} and \emph{knowledge-based}  social authentications. In trustee-based social authentications~\cite{Brainard06},  the selected friends aid the user in the authentication process. Knowledge-based social authentication, however, asks the user questions about his or her selected friends, and thus friends are not directly involved.

\myparatight{Trustee-based social authentication systems} Authentication is traditionally based on three factors: \emph{something you know} (e.g., a password), \emph{something you have} (e.g., a RSA SecurID), and \emph{something you are} (e.g., fingerprint). 

Brainard et al.~\cite{Brainard06} proposed to use the fourth factor, i.e., \emph{somebody you know}, to authenticate users. We call the fourth factor  trustee-based social authentication. Originally, Brainard et al. combined trustee-based social authentication with some other factor as a two-factor authentication mechanism. It was later adapted to be a backup authenticator~\cite{facebookTF,facebookSocAuth,Schechter09}. For instance, Schechter et al. [2] designed and built a prototype of trustee-based social authentication system which was integrated into Microsoft’s Windows Live ID system. Moreover, Facebook designed Trusted Friends in October, 2011~\cite{facebookTF}, and it was improved to be Trusted Contacts~\cite{facebookSocAuth} in May, 2013.

\myparatight{Knowledge-based social authentication systems} Such social authentications are still based on \emph{something you know}. Yardi et al.~\cite{Yardi08} proposed a knowledge-based authentication system based on photos to test if a user belongs to the group (e.g., interest groups in Facebook) that he or she tries to access. Facebook recently launched a similar photo-based social authentication system~\cite{Rice11}, in which Facebook shows a few photos of a friend of a user and asks the user to name the friend. Such system essentially relies on the knowledge that the user knows the person in the shown photos. However, recent work has shown, via theoretical modeling~\cite{Kim12-soc-auth} and empirical evaluations~\cite{Polakis12}, that photo-based social authentications are not resilient to various attacks such as automatic face recognition techniques, questioning their use as a backup authentication mechanism. 


\subsection{Diffusion models}
Our forest fire attacks essentially describe diffusion processes in a trustee network.  We review a few representative diffusion models from different research areas and discuss the differences between them and our work.

\myparatight{Updates propagation models} Malkhi et al.~\cite{updatespropagation} proposed the $l$-Tree propagation model to diffuse updates among a large distributed system of data replicas, some of which might exhibit Byzantine failures. Their model assumes a point-to-point communication for each pair of nodes.  A node that already receives the update is called \emph{active}, otherwise it is called \emph{inactive}. Initially, a small set of nodes are active. Each active node is associated with a candidate set of nodes. In each iteration, each active node is allowed to send the update to at most $F$ nodes which are selected from the corresponding candidate set uniformly at random. An inactive node becomes active if it receives the update from at least $k$ other nodes. 

There are two key differences between our forest fire attacks and the $l$-Tree propagation model. First, an uncompromised (i.e., inactive) node can receive verification codes (i.e., updates) from uncompromised trustees via spoofing attacks in forest fire attacks while an inactive node can only receive updates from active nodes in the $l$-Tree model. Second, in each iteration, each compromised node sends verification codes to \emph{all} nodes that select it as a trustee in forest fire attacks while an active node can only send the update to  at most $F$ nodes in the $l$-Tree model.

\myparatight{Information propagation models}
Models for how products and  innovations propagate on a social network have been studied in various domains such as viral marketing and  the spread of technological innovations. These models can be divided into two categories~\cite{Kempe03}: \emph{linear threshold model} and \emph{independent cascade model}. Again, a node is said to be \emph{active} if it already adopts the corresponding product or innovation, otherwise it is \emph{inactive}. In both models, a small set of  nodes are active initially.

\emph{Linear threshold model:} In this model, each node $u$ is associated with a threshold,  which indicates the fraction of $u$'s friends that must become active before $u$ becomes active.  The propagation proceeds deterministically in discrete iterations: in the $t$th iteration, an inactive node becomes active if its fraction of active friends is no less than $u$' activation threshold.

\emph{Independent cascade model:} Different  from the linear threshold model, the independent cascade model proceeds in discrete iterations according to the following randomized rule: when a node $u$ first becomes active in the $t$th iteration, it has a \emph{single} chance to activate each of its currently inactive friend $v$ and succeed with some probability which encodes the influence of $u$ to $v$.

The key difference is that verification codes can be propagated from an uncompromised (i.e., inactive) node to another uncompromised node via spoofing attacks in forest fire attacks while an inactive node can only be activated by active nodes in the linear threshold model and the independent cascade model.  Moreover, an active node only has a single chance to activate its friends in the independent cascade model while a compromised node can send verification codes to the uncompromised nodes that select it as a trustee as many times as it wants. 


\myparatight{Epidemic propagation models} Epidemic propagation models describe the propagations of various contagious diseases such as sexually transmitted diseases, inﬂuenza, and measles.  

One popular epidemic model is the so-called SIR model (Chapter 21 of~\cite{Easley10}). Different from the above reviewed models in which each node can be either active or inactive, SIR model assumes that each node can be in one of the three states, i.e., \emph{susceptible}, \emph{infectious}, and \emph{removed}.  Initially, a set of nodes are infectious and all other nodes are susceptible. Each infectious node $u$ remains infectious for a fixed number of iterations $I$. In each of the $I$ iterations, $u$ has some probability of passing the disease to each of its susceptible neighbors.  After $I$ iterations, $u$ becomes removed, which means that $u$ cannot catch nor propagate the disease any more.

Again, a susceptible (i.e., uncompromised) node can only be influenced by  infectious (i.e., compromised) nodes in the SIR model, which is different from the forest fire attacks. Moreover, the state \emph{removed} is not meaningful in the context of forest fire attacks since a node could be compromised again even if it recovers the account and resets the password. 

\myparatight{Summary} The key difference between forest fire attacks and previous diffusion models lies in the spoofing attacks. Specifically, an uncompromised node can obtain verification codes from its uncompromised trustees via spoofing attacks in the forest fire attacks while an inactive or susceptible node can only be influenced by its active or infectious neighbors in previous diffusion models. 

\section{Conclusion and Future Work}
In this paper, we provide the first systematic study about the security of trustee-based social authentications. First, we introduce \emph{forest fire attacks}. In these attacks, an attacker first obtains a small number of compromised seed users and then iteratively attacks the rest of users according to a priority ordering of them. Second, we construct a probabilistic model to formalize the threats of forest fire attacks and their costs for attackers. Third, we introduce a few strategies to select seed users and construct priority orderings, and we discuss various defense strategies. Fourth, via extensive evaluations using three real-world social network datasets, we find that forest fire attack is a potential big threat. For instance, with a small number (e.g., 1,000) of seed users, an attacker can further compromise two to three orders of magnitude more users in some scenarios with low (or even no) costs of sending spoofing messages. However, our defense strategy, which guarantees that no users are trustees of too many other users, can decrease the number of compromised users by one to two orders of magnitude and increase the costs for attackers by a few times in some cases. Moreover,  the recovery threshold should be set to be 4 to better balance between security and usability.

A few future directions include evaluating forest firest attacks on real social authentication systems such as Facebook's Trusted Contacts, designing new attack and defense strategies, and optimizing forest fire attacks given a time constraint.


%

\appendices
\section{The attack maximization problem is NP-Complete}
\label{theorem:attack:proof}
To prove the hardness of the attack maximization problem, we only need to show the hardness of a corresponding decision problem, i.e., given $G_T,k,n_s,n,$ and an integer $l$, it is NP-Complete to decide whether $n_c(G_T,k,n_s,n)\geq l$. This follows from a  reduction from the NP-Complete Set Cover problem: given a ground set $X=\{x_1,\ldots,x_a\}$, and a collection of its subsets $S_1, S_2,\cdots, S_m$, we want to determine if there exists $t$ subsets $S_{i_1},S_{i_2},\cdots,S_{i_t}$ such that $S_{i_1}\cup S_{i_2} \cup \cdots \cup S_{i_t}=X$. We show that this can be viewed as a special case of the attack maximization problem, where the spoofing probabilities and recovery probabilities are all $0$. 

Given any instance of a set cover problem, we construct a corresponding trustee network as follows: there are $k$ nodes for each $S_j$, and one node for each $x_i$; each $x_i$ has all the $k$ copies of $S_j$ as trustees iff $x_i\in S_j$; in addition, for each $S_j$ we have $a^2k^2$ dummy nodes each of which designates the $k$ copies of $S_j$ as trustees. Then the Set Cover problem is satisfiable iff the attacker can compromise at least $l=ta^2k^2+tk+a$ nodes with a seed set whose size is $tk$, i.e., $n_c(G_T,k,tk,n)\geq l$.

\end{document}